\documentclass[10pt,twocolumn,twoside]{IEEEtran}
\usepackage{calc}

\usepackage{multirow}
\usepackage{cite}
\usepackage{graphicx,subfigure}
\usepackage{psfrag}
\usepackage{amsmath,amssymb}
\usepackage{color}

\newcommand{\diag}{\mathrm{diag}}

\DeclareMathOperator*{\dotleq}{\overset{.}{\leq}}
\DeclareMathOperator*{\dotgeq}{\overset{.}{\geq}}
\DeclareMathOperator*{\defeq}{\triangleq}

\newtheorem{theorem}{Theorem}
\newtheorem{corollary}{Corollary}[theorem]

\newtheorem{lemma}{Lemma}

\newtheorem{proposition}{Proposition}
\newtheorem{example}{Example}

\def\bfg{{\boldsymbol{g}}}

\def\bfh{{\boldsymbol{h}}}

\newcommand{\bit}{\begin{itemize}}
\newcommand{\eit}{\end{itemize}}

\newcommand{\bc}{\begin{center}}
\newcommand{\ec}{\end{center}}

\newcommand{\ba}{\begin{array}}
\newcommand{\ea}{\end{array}}

\newcommand{\beq}{\begin{equation}}
\newcommand{\eeq}{\end{equation}}

\newcommand{\beqn}{\begin{equation*}}
\newcommand{\eeqn}{\end{equation*}}

\newcommand{\bean}{\begin{eqnarray*}}
\newcommand{\eean}{\end{eqnarray*}}
\newcommand{\bea}{\begin{eqnarray}}
\newcommand{\eea}{\end{eqnarray}}

\def\C{\mathbb{C}}
\def\E{\mathbb{E}}

\def\cv{\boldsymbol{c}}

\def\ev{\boldsymbol{e}}

\def\gv{\boldsymbol{g}}
\def\hv{\boldsymbol{h}}

\def\qv{\boldsymbol{q}}

\def\wv{\boldsymbol{w}}
\def\xv{\boldsymbol{x}}
\def\yv{\boldsymbol{y}}
\def\zv{\boldsymbol{z}}

\def\Mm{\boldsymbol{M}}

\def\Sm{\boldsymbol{S}}

\newcommand{\Bc}{{\mathcal B}}

\newcommand{\Xc}{{\mathcal X}}

\newcommand{\T}{{\scriptscriptstyle\mathsf{T}}}
\renewcommand{\H}{{\scriptscriptstyle\mathsf{H}}}

\newtheorem{remark}{Remark}

\newcommand{\Psim}{\pmb{\Psi}}
\renewcommand{\Bmatrix}[1]{\begin{bmatrix}#1\end{bmatrix}}

\begin{document}
\sloppy

\title{Toward the Performance vs. Feedback Tradeoff \\ for the Two-User MISO Broadcast Channel}
\author{Jinyuan Chen and Petros Elia
\thanks{The research leading to these results has received funding from the European Research Council under the European Community's Seventh Framework Programme (FP7/2007-2013) / grant agreement no. 257616 (CONECT), from the FP7 CELTIC SPECTRA project, and from Agence Nationale de la Recherche project ANR-IMAGENET.
}
\thanks{J. Chen and P. Elia are with the Mobile Communications Department, EURECOM, Sophia Antipolis, France (email: \{chenji, elia\}@eurecom.fr)}
\thanks{This paper was presented in part at the 50th Annual Allerton Conference 2012, at ITA-UCSD 2013, and at ISIT 2013.}
\thanks{Copyright (c) 2013 IEEE. Personal use of this material is permitted.  However, permission to use this material for any other purposes must be obtained from the IEEE by sending a request to pubs-permissions@ieee.org.}
}

\maketitle
\thispagestyle{empty}

\begin{abstract}
For the two-user MISO broadcast channel with imperfect and delayed channel state information at the transmitter (CSIT), the work explores the tradeoff between \emph{performance} on the one hand, and \emph{CSIT timeliness and accuracy} on the other hand. The work considers a broad setting where communication takes place in the presence of a random fading process, and in the presence of a feedback process that, at any point in time, may provide CSIT estimates - of some arbitrary accuracy - for any past, current or future channel realization. This feedback quality may fluctuate in time across all ranges of CSIT accuracy and timeliness, ranging from perfectly accurate and instantaneously available estimates, to delayed estimates of minimal accuracy. Under standard assumptions, the work derives the degrees-of-freedom (DoF) region, which is tight for a large range of CSIT quality. This derived DoF region concisely captures the effect of channel correlations, the accuracy of predicted, current, and delayed-CSIT, and generally captures the effect of the quality of CSIT offered at any time, about any channel.

The work also introduces novel schemes which - in the context of imperfect and delayed CSIT - employ encoding and decoding with a phase-Markov structure.
The results hold for a large class of block and non-block fading channel models, and they unify and extend many prior attempts to capture the effect of imperfect and delayed feedback.  This generality also allows for consideration of novel pertinent settings, such as the new \emph{periodically evolving feedback} setting, where a gradual accumulation of feedback bits progressively improves CSIT as time progresses across a finite coherence period.
\end{abstract}


\section{Introduction}
\subsection{Channel model}
We consider the multiple-input single-output broadcast channel (MISO BC) with an $M$-transmit antenna ($M\geq 2$) transmitter communicating to two receiving users with a single receiving antenna each.  Let $\hv_{t},\gv_{t}$ denote the channel of the first and second user respectively at time $t$, and let $\xv_{t}$ denote the transmitted vector at time $t$, satisfying a power constraint $\E[ ||\xv_{t}||^2 ] \le P$, for some power $P$ which also here takes the role of the signal-to-noise ratio (SNR). Here $\hv_t$ and $\gv_t$ are drawn from a random distribution, such that each has zero mean and identity covariance (spatially - but not necessarily temporally - uncorrelated), and such that $\hv_t$ is linearly independent of $\gv_t$ with probability 1.

In this setting, the corresponding received signals at the first and second user take the form
\begin{align}
y^{(1)}_t &= \hv^{\T}_t \xv_t + z^{(1)}_t      \label{eq:blockfy1}\\
y^{(2)}_t &= \gv^{\T}_t \xv_t + z^{(2)}_t      \label{eq:blockfy2}
\end{align}
($t=1,2,\cdots$), where $z^{(1)}_t,z^{(2)}_t$ denote the unit power AWGN noise at the receivers.

In the high-SNR setting of interest, for an achievable rate pair $(R_1,R_2)$ for the first and second user respectively, the corresponding degrees-of-freedom (DoF) pair $(d_1,d_2)$ is given by \[d_i = \lim_{P \to \infty} \frac{R_i}{\log P},\ i=1,2\] and the corresponding DoF region is then the set of all achievable DoF pairs.

\subsection{Delay-and-quality effects of feedback}
As in many multiuser wireless communications scenarios, the performance of the broadcast channel depends on the timeliness and precision of channel state information at the transmitter (CSIT).  This timeliness and precision though may be reduced by limited-capacity feedback links, which may offer CSIT with consistently low precision and high delays, i.e., feedback that offers an inaccurate representation of the true state of the channel, as well feedback that can only be used for an insufficient fraction of the communication duration. The corresponding performance degradation, as compared to the case of having perfect feedback without delay, forces the delay-and-quality question of how much CSIT precision is necessary, and when, in order to achieve a certain performance.

\subsection{Channel process and feedback process with predicted, current, and delayed CSIT}
We here consider communication of an infinite duration $n$, a channel fading process $\{\bfh_t,\bfg_t\}_{t=1}^n$ drawn from a statistical distribution, and a feedback process that provides CSIT estimates $$\{\hat{\bfh}_{t,t'},\hat{\bfg}_{t,t'}\}_{t,t' = 1}^n$$ (of channel $\bfh_{t},\bfg_{t}$) at any time $t'$ - before, during, or after materialization of $\bfh_t,\bfg_t$ at time $t$ - and does so with precision/quality defined by the statistics of
\beq \label{eq:feedbackQuality} \{(\bfh_t-\hat{\bfh}_{t,t'}),(\bfg_t-\hat{\bfg}_{t,t'})\}_{t,t' = 1}^n\eeq
where we consider these estimation errors to have zero-mean circularly-symmetric complex Gaussian entries.
Naturally any attempt to capture and meet the tradeoff between performance, and feedback timeliness and quality, must consider the full effect of the statistics of the channel and of CSIT precision $\{(\bfh_t-\hat{\bfh}_{t,t'}),(\bfg_t-\hat{\bfg}_{t,t'})\}_{t,t' = 1}^n$ at any point in time, about any channel.

\subsubsection{Predicted, current, and delayed CSIT}
For the channel $\bfh_t,\bfg_t$ at time $t$, the set of all estimates $\{\hat{\bfh}_{t,t'},\hat{\bfg}_{t,t'}\}_{t' = 1}^n$ is formed by what can be described as the set of \emph{predicted estimates} $\{\hat{\bfh}_{t,t'},\hat{\bfg}_{t,t'}\}_{t'<t}$, by the \emph{current estimates} $\hat{\bfh}_{t,t},\hat{\bfg}_{t,t}$ at time $t$, and by the set of \emph{delayed CSIT} $\{\hat{\bfh}_{t,t'},\hat{\bfg}_{t,t'}\}_{t'>t}$ comprising of estimates that are not available at time $t$. Predicted CSIT may potentially allow for reduction of the effect of future interference, current CSIT may be used to `separate' the current signals of the users, while delayed CSIT may facilitate retrospective compensation for the lack of perfect quality feedback (\cite{MAT:11c}).


\subsection{Notation, conventions and assumptions}
We will use the notation
\beq \label{eq:alpha1} \alpha_{t}^{(1)}\defeq  -\lim_{P\rightarrow \infty}  \frac{ \log \E[||\bfh_t - \hat{\bfh}_{t,t}||^2]  }{  \log P } \eeq
\beq \label{eq:alpha2} \alpha_{t}^{(2)}\defeq  -\lim_{P\rightarrow \infty}  \frac{ \log\E[||\bfg_t - \hat{\bfg}_{t,t}||^2]  }{  \log P }\eeq
to describe the \emph{current quality exponent} for the current estimate of the channel of each user at time $t$ ($\alpha_{t}^{(1)}$ is for user~1), while we will use
\beq \label{eq:beta1} \beta_{t}^{(1)}\defeq  -\lim_{P\rightarrow \infty}  \frac{ \log\E[||\bfh_t - \hat{\bfh}_{t,t+\eta}||^2]  }{  \log P } \eeq
\beq \label{eq:beta2} \beta_{t}^{(2)}\defeq  -\lim_{P\rightarrow \infty}  \frac{\log \E[||\bfg_t - \hat{\bfg}_{t,t+\eta}||^2]  }{  \log P }\eeq
- for any sufficiently large but finite integer $\eta>0$ - to denote the \emph{delayed quality exponent} for each user. To clarify, with delayed CSIT consisting of all channel estimates that arrive after the channel materializes, the above use of a finite $\eta$, reflects the fact that we here only consider delayed CSIT that arrives up to a finite time of $\eta$ channel uses from the moment the channel materializes.
In words, $\alpha_{t}^{(1)}$ measures the precision/quality of the CSIT (about $\bfh_t$) that is available at time $t$, while $\beta_{t}^{(1)}$ measures the (best) quality of the CSIT (again about $\bfh_t$) which arrives strictly after the channel appears, i.e., strictly after time $t$ (similarly $\alpha_{t}^{(2)},\beta_{t}^{(2)}$ for the channel $\bfg_t$ of the second user).

It is easy to see that without loss of generality, in the DoF setting of interest, we can restrict our attention to the range~\footnote{To see this, we recall from \cite{Jindal:06m,Caire+:10m} that under a peak-power constraint of $P$, having CSIT estimation error in the order of $P^{-1}$ causes no DoF reduction as compared to the perfect CSIT case.  In our DoF high-SNR setting of interest where $P>>n$, this same observation also holds under an average power constraint of $P$. The fact that $\alpha_t^{(i)} \leq \beta_t^{(i)}$ comes naturally from the fact that one can recall, at a later time, statistically good estimates.} \beq \label{eq:exponentRange} 0\leq \alpha_t^{(i)} \leq \beta_t^{(i)} \leq 1\eeq
where $\beta_t^{(1)} = \beta_t^{(2)} = 1$ corresponds to being able to eventually gather (asymptotically) perfect delayed CSIT for $\bfh_t,\bfg_t$, while $\alpha_t^{(1)} = \alpha_t^{(2)} = 1$, simply corresponds to having instantaneously available CSIT of asymptotically perfect precision.

Furthermore we will use the notation
\beq \label{eq:averages}
\bar{\alpha}^{(i)} \defeq  \lim_{n \rightarrow \infty}\frac{1}{n}\sum^{n}_{t=1}\alpha_t^{(i)}, \quad \bar{\beta}^{(i)} \defeq \lim_{n \rightarrow \infty}\frac{1}{n}\sum^{n}_{t=1}\beta_t^{(i)}, \quad i=1,2\eeq
to denote the average of the quality exponents.

Throughout this paper, $(\bullet)^\T$, $(\bullet)^{\H}$ and $||\bullet||_{F}$ will denote the transpose, conjugate transpose and Frobenius norm of a matrix respectively, while $\diag(\bullet)$ will denote a diagonal matrix, $||\bullet||$ will denote the Euclidean norm, and $|\bullet|$ will denote the magnitude of a scalar. $o(\bullet)$ comes from the standard Landau notation, where $f(x) = o(g(x))$ implies $\lim_{x\to \infty} f(x)/g(x)=0$.  We will also use $\doteq$ to denote \emph{exponential equality}, i.e., we write $f(P)\doteq P^{B}$ to denote $\displaystyle\lim_{P\to\infty}\frac{\log f(P)}{\log P}=B$.  Similarly $\dotgeq$ and $\dotleq$ will denote exponential inequalities.  Logarithms are of base~$2$.

\subsubsection{Assumptions}
Our results, specifically the achievability part, will hold under the soft assumption that any sufficiently long subsequence $\{\alpha^{(1)}_t\}_{t=\tau}^{\tau+T}$ (resp. $\{\alpha^{(2)}_t\}_{t=\tau}^{\tau+T},\{\beta^{(1)}_t\}_{t=\tau}^{\tau+T},\{\beta^{(2)}_t\}_{t=\tau}^{\tau+T}$) has an average that approaches the long term average $\bar{\alpha}^{(1)}$ (resp. $\bar{\alpha}^{(2)},\bar{\beta}^{(1)},\bar{\beta}^{(2)}$), for some finite $T$ that can be chosen to be sufficiently large to allow for the above convergence. Such an assumption - which has also been employed in works like~\cite{TJSP:12} - essentially imply that the long term statistics of the feedback process, remains the same in time, i.e., that the average feedback behavior - averaged over large amounts of time - remains the same throughout the communication process.

We also adhere to the common convention (see~\cite{MAT:11c,YKGY:12d,GJ:12o,MJS:12}) of assuming perfect and global knowledge of channel state information at the receivers (perfect global CSIR), where the receivers know all channel states and all estimates.
We further adopt the common assumption (see \cite{LSW:05,KYGY:12o,YKGY:12d,GJ:12o}) that the current \emph{estimation error} is statistically independent of current and past \emph{estimates}, and consequently that the input signal is a function of the message and of the CSIT.  This assumption fits well with many channel models spanning from the fast fading channel (i.i.d. in time), to the correlated channel model as this is considered in \cite{KYGY:12o}, to the quasi-static block fading model where the CSIT estimates are successively refined while the channel remains static (see~\cite{Jindal:06m}, see also the discussion in the appendix in Section~\ref{sec:DiscussionAssumption}). Additionally we consider the entries of each estimation error vector $\bfh_t - \hat{\bfh}_{t,t'}$ (similarly of $\bfg_t - \hat{\bfg}_{t,t'}$) to be i.i.d. Gaussian, clarifying though that we are just referring to the $M$ entries in each such specific vector $\bfh_t - \hat{\bfh}_{t,t'}$, and that we do \emph{not} suggest that the error entries are i.i.d. in time or across users. The appendix in Section~\ref{sec:DiscussionAssumption} offers further details and justification on the above assumptions and conventions.

Finally we safely assume that $\E[||\bfh_t - \hat{\bfh}_{t,t'}||^2]   \ \dotleq \ \E[||\bfh_t - \hat{\bfh}_{t,t''}||^2] $ (similarly $\E[||\bfg_t - \hat{\bfg}_{t,t'}||^2] \ \dotleq \   \E[||\bfg_t - \hat{\bfg}_{t,t''}||^2] $), for any $t'>t''$.  This assumption - which simply suggests that one can revert back to past estimates of statistically better quality - is used here for simplicity of notation, and can be removed, after a small change in the definition of the quality exponents, without an effect to the main result.

\subsection{Prior work}
The delay-and-quality effects of feedback, naturally fall between the two extreme cases of no CSIT and of full CSIT (immediately available, perfect-quality CSIT), with full CSIT allowing for the optimal $1$ DoF per user (cf.~\cite{CS:03}), while the absence of any CSIT reduces this to just $1/2$ DoF per user (cf.~\cite{JG:05,HJSV:12}).

Toward bridging this gap, different works have considered the use of imperfect and delayed feedback.  For example, the work by Lapidoth, Shamai and Wigger in \cite{LSW:05} considered the case where the amount of feedback is limited to the extent that the channel-estimation error power does not vanish with increasing SNR, in the sense that $\lim_{P \rightarrow \infty} (\log\E[||\bfh_t-\hat{\bfh}_{t,t}||^2])/\log P = \lim_{P \rightarrow \infty} (\log\E[||\bfg_t-\hat{\bfg}_{t,t}||^2])/\log P = 0$.  In this setting - which corresponds to the case here where $\alpha^{(1)}_t=\alpha^{(2)}_t=\beta^{(1)}_t=\beta^{(2)}_t = 0, \ \forall t$ - the work in \cite{LSW:05} showed that the symmetric DoF is upper bounded by $2/3$ DoF per user, again under the assumption placed here that the input signaling is independent of the estimation error. It is worth noting that finding the exact DoF in this zero-exponent setting, currently remains an open problem.

At the other extreme, the work by Caire et al.~\cite{Caire+:10m} (see also the work of Jindal~\cite{Jindal:06m}, as well as of Lapidoth and Shamai~\cite{LS:02}) showed that having immediately available CSIT estimates with estimation error power that is in the order of $P^{-1}$ - i.e., having $-\lim_{P \rightarrow \infty} (\log\E[||\bfh_t-\hat{\bfh}_{t,t}||^2])/\log P = -\lim_{P \rightarrow \infty} (\log\E[||\bfg_t-\hat{\bfg}_{t,t}||^2])/\log P = 1, \ \forall t$, corresponding here to having $\alpha^{(1)}_t=\alpha^{(2)}_t=1, \ \forall t$ - causes no DoF reduction as compared to the perfect CSIT case, and can thus achieve the optimal $1$ DoF per user.

A valuable tool toward bridging this gap and further understanding the delay-and-quality effects of feedback, came with the work by Maddah-Ali and Tse in~\cite{MAT:11c} which showed that arbitrarily delayed feedback can still allow for performance improvement over the no-CSIT case. In a fast-fading block-fading setting, the work differentiated between current and delayed CSIT - with delayed CSIT defined in~\cite{MAT:11c} as the CSIT which is available after the channel's coherence period - and showed that delayed and completely obsolete CSIT, even without any current CSIT, allows for an improved optimal $2/3$ DoF per user. A key ingredient in employing such delayed CSIT, was found in a form of retrospective interference alignment. This setting - which in principle corresponded to perfect delayed CSIT - is here represented by current-CSIT exponents of the form $\alpha^{(1)}_t=\alpha^{(2)}_t= 0, \ \forall t$.

Within the same block-fading context of delayed vs. current CSIT, the work by Kobayashi et al., Yang et al., and Gou and Jafar~\cite{KYGY:12o,YKGY:12d,GJ:12o}, quantified the usefulness of combining delayed and completely obsolete CSIT with immediately available but imperfect CSIT of a certain quality $\alpha= -\lim_{P \rightarrow \infty} (\log\E[||\bfh_t-\hat{\bfh}_{t,t}||^2])/\log P = -\lim_{P \rightarrow \infty} (\log\E[||\bfg_t-\hat{\bfg}_{t,t}||^2])/\log P$ that remained unchanged throughout the communication process. This work - which again in principle assumed perfect delayed CSIT, and which is here represented by current-CSIT exponents of the form $\alpha^{(1)}_t=\alpha^{(2)}_t= \alpha, \ \forall t$ - derived the optimal DoF region to be that with a symmetric DoF of $(2+\alpha)/3$ DoF per user. A key enabling ingredient here was introduced in \cite{KYGY:12o} in the form of interference quantizing and forwarding.

Interestingly, despite the fact that in principle, the above settings in \cite{MAT:11c,KYGY:12o,YKGY:12d,GJ:12o} corresponded to perfect delayed CSIT, the actual schemes in these works in fact achieved the optimal DoF, by using delayed CSIT for only a fraction of the channels.  This possibility that imperfect and sparse delayed CSIT may be as good as perfect and omnipresent delayed CSIT (cf.~\cite{CE:12c}), is one of the many facets that are explored in detail in Sections~\ref{sec:bc-dof}-\ref{sec:evolvingCSIT}.

Another interesting approach was introduced by Tandon et al. in~\cite{TJSP:12} who considered the fast-fading two-user MISO BC setting, where each user's CSIT changes every coherence period by alternating between the three extreme states of perfect current CSIT, perfect delayed CSIT, and no CSIT.

Additionally, Lee and Heath in~\cite{LH:12} considered, in the setting of the quasi-static block-fading channel, the possibility that current CSIT may be available only after a certain fraction $\gamma$ of a finite-duration coherence period $T_c$.

Other work such as that by Maleki et al. in~\cite{MJS:12} considered, again in the MISO BC context, an asymmetric setting where both users offered perfect delayed CSIT, but where only one user offered perfect current CSIT while the other user offered no current CSIT.
In this setting, the optimal DoF corner point was calculated to be $(1,1/2)$ (sum-DoF $d_1+d_2 = 3/2$).
Another asymmetric-feedback setting was considered in~\cite{CE:12d}.

Other related works can be found in~\cite{GMK:11o,AGK:11o,GMK:11i,XAJ:11b,LSW:12,CE:13isit,CYE:13isit,KYG:13,HC:13,VMA:13,CE:13spawc,LHA:13,LSY:13,ZS:13,AA:13,GNS:07}, including interesting extensions in \cite{VV:11t} to the multiple-input multiple-output (MIMO) BC setting with no current CSIT, and recent work in \cite{YYGK:12} with extensions to the MIMO BC and MIMO IC (interference channel) setting with fixed-quality current CSIT.


\subsection{Structure of paper}

Section~\ref{sec:bc-dof} will give the main result of this work by describing, under the aforementioned common assumptions, the DoF offered by a CSIT process $\{\hat{\bfh}_{t,t'},\hat{\bfg}_{t,t'}\}_{t=1,t'=1}^n$ of a certain quality
$\{(\bfh_t-\hat{\bfh}_{t,t'}),(\bfg_t-\hat{\bfg}_{t,t'})\}_{t=1,t'=1}^n$.  Specifically Proposition~\ref{prop:genCSITInner} and Lemma~\ref{lem:bc-evol-outerb} lower and upper bound the DoF region, and the resulting Theorem~\ref{thm:genCSIT} provides the optimal DoF for a large range of `sufficiently good' delayed CSIT. The results capture specific existing cases of interest, such as the Maddah-Ali and Tse setting in~\cite{MAT:11c}, the Yang et al. and Gou and Jafar setting in~\cite{YKGY:12d,GJ:12o}, the Lee and Heath `not-so-delayed CSIT' setting in~\cite{LH:12} for two users, the Maleki et al. asymmetric setting in~\cite{MJS:12}, and in the range of sufficiently good delayed CSIT, also capture the results in the Tandon et al. setting of alternating CSIT~\cite{TJSP:12} .

Towards gaining further insight, we then proceed to provide different corollaries for specific cases of interest. Again in Section~\ref{sec:bc-dof}, Corollary~\ref{cor:sym} distills the main result down to the symmetric feedback case where $\bar{\alpha}^{(1)}= \bar{\alpha}^{(2)}$ and $\bar{\beta}^{(1)}= \bar{\beta}^{(2)}$, and immediately after that, Corollary~\ref{cor:Asyvssym} explores the benefits of such feedback symmetry, by quantifying the extent to which having similar feedback quality for the two users, offers a gain over the asymmetric case where one user has generally more feedback than the other. One of the outcomes here is that such `symmetry gains' are often nonexistent.
Corollary~\ref{cor:MalekiGeneralization} generalizes the pertinent result in the setting in ~\cite{MJS:12} corresponding to feedback asymmetry; a setting which we consider to be important as it captures the inherent non-homogeneity of feedback quality of different users.
Corollary~\ref{cor:ImperfectVsPerfectDelayed} offers insight on the need for delayed CSIT, and shows how, reducing $\bar{\alpha}^{(1)},\bar{\alpha}^{(2)}$ allows - to a certain extent - for further reducing of $\bar{\beta}^{(1)},\bar{\beta}^{(2)}$, without an additional DoF penalty. It will be surprising to note that the expressions from Corollary~\ref{cor:ImperfectVsPerfectDelayed}, match the amount of delayed CSIT used by different previous schemes which were designed for settings that in principle offered perfect delayed CSIT, and which were thus designed without an expressed purpose of reducing the amount of delayed CSIT.
At the other extreme, Corollary~\ref{cor:Prediction} offers insight on the need for using predicted channel estimates (forecasting channel states in advance), by showing that - at least in the range of sufficiently good delayed CSIT - employing predicted CSIT is unnecessary.

Section~\ref{sec:evolvingCSIT} highlights the newly considered \emph{periodically evolving feedback} setting over the quasi-static block fading channel, where a gradual accumulation of feedback, results in a progressively increasing CSIT quality as time progresses across a finite coherence period.
This setting is powerful as it captures the many feedback options that one may have in a block-fading environment where the \emph{statistical} nature of feedback may remain largely unchanged across coherence periods.
To offer further understanding, we provide examples which - under very clearly specified assumptions - describe how many feedback bits to introduce, and when, in order to achieve a certain DoF performance. In the same section, smaller results and examples offer further insight - again in the context of periodically evolving feedback over a quasi-static channel - like for example the result in Corollary~\ref{cor:HowMuchAndWHenForDOF}
which bounds the quality of current and of delayed CSIT needed to achieve a certain target symmetric DoF, and in the process offers intuition on when delayed feedback is entirely unnecessary, in the sense that there is no need to wait for feedback that arrives after the end of the coherence period of the channel.
Similarly Corollary~\ref{cor:delayWithConstraints} provides insight on the feedback delays that allow for a given target symmetric DoF in the presence of constraints on current and delayed CSIT qualities. This quantifies to a certain extent the intuitive argument that, with a target DoF in mind, feedback delays must be compensated for, with high quality feedback estimates.

Section~\ref{sec:schemes} corresponds to the achievability part of the proof of the main result, and presents the general communication scheme that utilizes the available information of a CSIT process $\{\hat{\bfh}_{t,t'},\hat{\bfg}_{t,t'}\}_{t=1,t'=1}^n$, to achieve the corresponding DoF corner points.  This is done - by properly employing different combinations of zero forcing, superposition coding, interference compressing and broadcasting, as well as specifically tailored power and rate allocation - in order to transmit private information, using currently available CSIT estimates to reduce interference, and using delayed CSIT estimates to alleviate the effect of past interference.  The scheme has a forward-backward phase-Markov structure which, in the context of imperfect and delayed CSIT, was first introduced in~\cite{CE:12d,CE:12c} to consist of four main ingredients that include, block-Markov encoding, spatial precoding, interference quantization, and backward decoding.\\
After the description of the scheme in its general form, and the explicit description of how the scheme achieves the different DoF corner points, Section~\ref{sec:scheme_exam} provides example schemes - distilled from the general scheme - for specific settings such as the imperfect-delayed CSIT setting, the (extended) alternating CSIT setting of Tandon et al.~\cite{TJSP:12}, as well as discusses schemes with small delay.

Section~\ref{sec:conclu} offers concluding remarks, the appendix in Section~\ref{sec:outerb} provides the details of the outer bound, the appendix in Section~\ref{sec:DetailsX} offers details on the proofs, while the appendix in Section~\ref{sec:DiscussionAssumption} offers a discussion on some of the assumptions employed in this work.

\vspace{1pt}

In the end, the above results provide insight on pertinent questions such as:
\bit
\item What CSIT feedback precision should be provided, and when, in order to achieve a certain target DoF performance? (Theorem~\ref{thm:genCSIT})
\item When is delayed feedback unnecessary? (Corollary~\ref{cor:HowMuchAndWHenForDOF})
\item Is there any gain in early prediction of future channels? (Corollary~\ref{cor:Prediction})
\item What current-CSIT and delayed-CSIT qualities suffice to achieve a certain performance?  (Corollary~\ref{cor:HowMuchAndWHenForDOF})
\item Can delayed CSIT that is sparse and of imperfect-quality, achieve the same DoF performance that was previously attributed to sending perfect delayed CSIT? (Corollary~\ref{cor:ImperfectVsPerfectDelayed})
\item How much more valuable are feedback bits that are sent early, than those sent late? (Section~\ref{sec:evolvingCSIT})
\item In the quasi-static block-fading case, is it better to send less feedback early, or more feedback later? (Section~\ref{sec:evolvingCSIT})
\item What is the effect of having asymmetric feedback links, and when can we have a `symmetry gain'? (Corollary~\ref{cor:Asyvssym})
\eit


\section{DoF region of the MISO BC\label{sec:bc-dof}}

We proceed with the main DoF results, which are proved in Section~\ref{sec:schemes} (inner bound) and Section~\ref{sec:outerb} (outer bound).

We here remind the reader of the sequences $\{\alpha_{t}^{(1)}\}_{t=1}^n,\{\alpha_{t}^{(2)}\}_{t=1}^n,\{\beta_{t}^{(1)}\}_{t=1}^n,\{\beta_{t}^{(2)}\}_{t=1}^n$ of quality exponents, as these were defined in \eqref{eq:alpha1}-\eqref{eq:beta2}, as well as of the corresponding averages $\bar{\alpha}^{(1)}, \bar{\alpha}^{(2)},\bar{\beta}^{(1)}, \bar{\beta}^{(2)}$ from~\eqref{eq:averages}. We also remind the reader that we consider communication over an asymptotically large time duration $n$.  We henceforth label the users so that $\bar{\alpha}^{(2)}\leq \bar{\alpha}^{(1)}$.

We start with the following proposition, the proof of which can be found in Section~\ref{sec:schemes} which describes the scheme that achieves the corresponding DoF corner points.

\begin{figure}
\centering
\includegraphics[width=9cm]{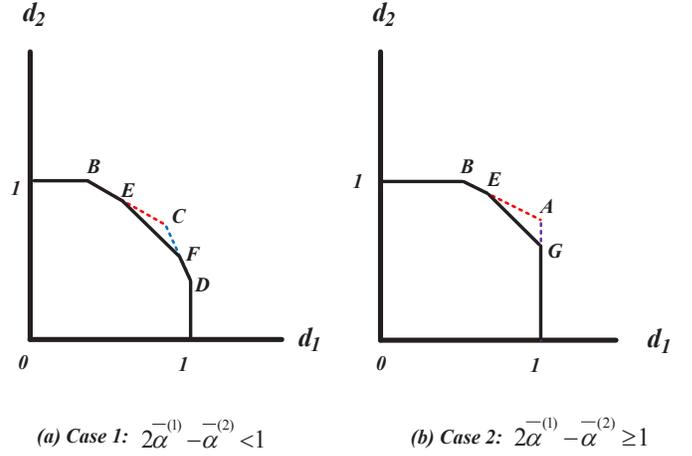}
\caption{ DoF region inner bound for the two-user MISO BC. The corner points take the following values: $E=(2\bar{\delta} - \bar{\alpha}^{(2)} , 1+\bar{\alpha}^{(2)}-\bar{\delta})$, $F=(1+\bar{\alpha}^{(1)}-\bar{\delta}, 2\bar{\delta} - \bar{\alpha}^{(1)})$, $G=(1, \bar{\delta})$, $B=(\bar{\alpha}^{(2)}, 1)$ and $D=(1, \bar{\alpha}^{(1)})$, where $\bar{\delta} \defeq	\min \{ \bar{\beta}^{(1)},  \bar{\beta}^{(2)}, \frac{ 1 + \bar{\alpha}^{(1)} +  \bar{\alpha}^{(2)}}{3},  \frac{1+  \bar{\alpha}^{(2)}}{2} \}$. }
\label{fig:DoFAsymmeticCSITImpDCSIT}
\end{figure}

\vspace{3pt}
\begin{proposition} \label{prop:genCSITInner}
The DoF region of the two-user MISO BC with a CSIT process $\{\hat{\bfh}_{t,t'},\hat{\bfg}_{t,t'}\}_{t=1,t'=1}^n$ of quality $\{(\bfh_t-\hat{\bfh}_{t,t'}),(\bfg_t-\hat{\bfg}_{t,t'})\}_{t=1,t'=1}^n$, is inner bounded by the polygon described by
  \begin{align}
        d_1 \leq  1, & \quad       d_2 \leq  1  \\
2 d_1 + d_2 &\leq  2 + \bar{\alpha}^{(1)}  \\
2 d_2 + d_1 &\leq  2 + \bar{\alpha}^{(2)}  \\
   d_1 +d_2 &\leq  1 + \min \{ \bar{\beta}^{(1)}, \bar{\beta}^{(2)}\}.
  \end{align}
\end{proposition}
\vspace{3pt}
Figure~\ref{fig:DoFAsymmeticCSITImpDCSIT} corresponds to the result in Proposition~\ref{prop:genCSITInner}.

Towards tightening the above inner bound, we here draw from the DoF outer bound in~\cite{YKGY:12d} that focused on CSIT with invariant and symmetric quality and non-static channels, and employ techniques that allow for the new bound to hold for a broad range of channels including the static-channel case that is of particular interest here. The proof of the new bound can be found in Section~\ref{sec:outerb}.

\vspace{3pt}
\begin{lemma} \label{lem:bc-evol-outerb}
The DoF region of the two-user MISO BC with a CSIT process $\{\hat{\bfh}_{t,t'},\hat{\bfg}_{t,t'}\}_{t=1,t'=1}^n$ of quality $\{(\bfh_t-\hat{\bfh}_{t,t'}),(\bfg_t-\hat{\bfg}_{t,t'})\}_{t=1,t'=1}^n$, is upper bounded as
  \begin{align} \label{eq:upperbound}
	   d_1 \le 1, &\quad  \  \  d_2 \le 1   \\
     2d_1  + d_2 &\le 2 +\bar{\alpha}^{(1)} \\
     2d_2  + d_1 &\le 2 +\bar{\alpha}^{(2)}.
  \end{align}
\end{lemma}
\vspace{3pt}

Comparing the above inner and outer bounds, and observing that the last bound in Proposition~\ref{prop:genCSITInner} becomes inactive in the range of sufficiently good delayed-CSIT where $\min \{ \bar{\beta}^{(1)}, \bar{\beta}^{(2)}\} \geq \min \{\frac{1+\bar{\alpha}^{(1)}+\bar{\alpha}^{(2)}}{3}, \frac{1+\bar{\alpha}^{(2)}}{2}\}$, gives the main result of this work in the form of the following theorem that provides the optimal DoF for this large range of `sufficiently good' delayed CSIT.
\vspace{3pt}
\begin{theorem} \label{thm:genCSIT}
The optimal DoF region of the two-user MISO BC with a CSIT process $\{\hat{\bfh}_{t,t'},\hat{\bfg}_{t,t'}\}_{t=1,t'=1}^n$ of quality $\{(\bfh_t-\hat{\bfh}_{t,t'}),(\bfg_t-\hat{\bfg}_{t,t'})\}_{t=1,t'=1}^n$ is given by
  \begin{align}
        d_1 \leq  1,  &\quad      d_2 \leq  1  \\
2 d_1 + d_2 &\leq  2 + \bar{\alpha}^{(1)}  \\
2 d_2 + d_1 &\leq  2 + \bar{\alpha}^{(2)}
  \end{align}
for any sufficiently good delayed-CSIT process such that $\min \{ \bar{\beta}^{(1)}, \bar{\beta}^{(2)}\} \geq \min \{\frac{1+\bar{\alpha}^{(1)}+\bar{\alpha}^{(2)}}{3}, \frac{1+\bar{\alpha}^{(2)}}{2}\}$.
\end{theorem}
\vspace{3pt}

As mentioned, the achievability part of the proof can be found in Section~\ref{sec:schemes}.

\begin{figure}
\centering
\includegraphics[width=9cm]{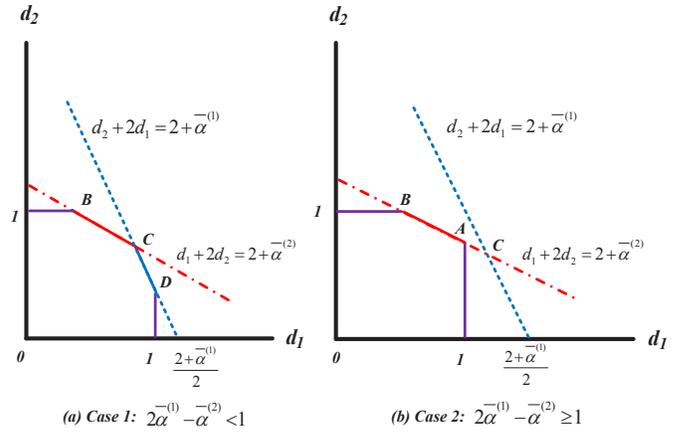}
\caption{Optimal DoF region for the two-user MISO BC, for the case of $\min \{ \bar{\beta}^{(1)}, \bar{\beta}^{(2)}\} \geq \min \{\frac{1+\bar{\alpha}^{(1)}+\bar{\alpha}^{(2)}}{3}, \frac{1+\bar{\alpha}^{(2)}}{2}\}$.
The corner points take the following values: $A=(1, \frac{1+\bar{\alpha}^{(2)}}{2})$, $B=(\bar{\alpha}^{(2)}, 1)$, $C=(\frac{2+2\bar{\alpha}^{(1)}-\bar{\alpha}^{(2)}}{3}, \frac{2+2\bar{\alpha}^{(2)}-\bar{\alpha}^{(1)}}{3})$, and $D=(1, \bar{\alpha}^{(1)})$. }
\label{fig:DoFAsymmeticCSIT}
\end{figure}

Figure~\ref{fig:DoFAsymmeticCSIT} corresponds to the main result in the theorem.

\vspace{2pt}

Before proceeding to specific corollaries that offer further insight, it is worth making a comment on the fact that the entire complexity of the problem is captured by the quality exponents.

\vspace{2pt}
\begin{remark}
The results suggest that the quality exponents capture - in the DoF setting of interest, and under our assumptions - the effect of the statistics of the CSIT precision $\{(\bfh_t-\hat{\bfh}_{t,t'}),(\bfg_t-\hat{\bfg}_{t,t'})\}_{t,t' = 1}^n$.
This is indeed the case since the following two hold.  Firstly, given the Gaussianity of the estimation errors,
the statistics of $\{(\bfh_t-\hat{\bfh}_{t,t'}),(\bfg_t-\hat{\bfg}_{t,t'})\}_{t,t' = 1}^n$ are captured by the $2n^2\times 2n^2$ covariance matrix\footnote{This size of the covariance matrix reflects the fact that the $M$ entries of each $\bfh_t-\hat{\bfh}_{t,t'}$ are i.i.d. (similarly of $\bfg_t-\hat{\bfg}_{t,t'}$). Please note that we refer to independence across the \emph{spatial} dimensions of the channel of one user, and certainly do not refer to independence across time or across users.} of the $2n^2$-length vector consisting of the elements $\{(\bfh_t-\hat{\bfh}_{t,t'}),(\bfg_t-\hat{\bfg}_{t,t'})\}_{t,t' = 1}^n$.  The diagonal entries of this covariance matrix are simply  $\{  \frac{1}{M}\E[||\bfh_t-\hat{\bfh}_{t,t'}||^2],  \frac{1}{M}\E[||\bfg_t-\hat{\bfg}_{t,t'}||^2]\}_{t,t' = 1}^n$.  With the above in mind, we also note that the outer bound has kept open the possibility of having arbitrary off-diagonal elements in this covariance matrix (this is specifically seen in the steps in~\eqref{eq:R12b},~\eqref{eq:R12b-opt}), thus allowing for the outer bound to hold irrespective of the off-diagonal elements of this covariance matrix. Consequently, under our assumptions, the essence of the statistics is captured by $\{\E[||\bfh_t-\hat{\bfh}_{t,t'}||^2],\E[||\bfg_t-\hat{\bfg}_{t,t'}||^2]\}_{t,t' = 1}^n$, and its effect is captured -  in the high-SNR DoF regime - by the quality exponents.
\end{remark}

\subsubsection{Symmetric vs. asymmetric feedback}
We proceed to explore the special case of symmetric feedback where the long-term accumulated feedback quality at the two users is similar, in the sense that the feedback links of user~1 and user~2 share the same long-term exponent \emph{averages} $\bar{\alpha}^{(1)}= \bar{\alpha}^{(2)}=:\bar{\alpha}$ and $\bar{\beta}^{(1)}= \bar{\beta}^{(2)}=:\bar{\beta}$. Most existing works, with an exception in \cite{MJS:12} and \cite{CE:12d}, fall under this symmetric feedback setting. The following holds directly from Theorem~\ref{thm:genCSIT} and Proposition~\ref{prop:genCSITInner}.

\vspace{3pt}
\begin{corollary}[DoF with symmetric feedback] \label{cor:sym}
The optimal DoF region for the symmetric feedback case, takes the form
\begin{align*}
	   d_1 \le 1, \quad  \  d_2 \le 1,  \quad   2d_1  + d_2 \le 2 +\bar{\alpha}, \quad     2d_2  + d_1 \le 2 +\bar{\alpha}
\end{align*}
when $\bar{\beta}\geq \frac{1+2\bar{\alpha}}{3}$, while when $\bar{\beta}<\frac{1+2\bar{\alpha}}{3}$ this region is inner bounded by the achievable region
  \begin{align} \label{eq:DoFEcsitImpD}
	   d_1 \le 1, \quad   d_2 \le 1    \\
     2d_1  + d_2 \le 2 +\bar{\alpha} \\
     2d_2  + d_1 \le 2 +\bar{\alpha} \\
		  d_2  + d_1 \le 1 +\bar{\beta}.
  \end{align}
\end{corollary}
\vspace{3pt}

\begin{figure}
\centering
\includegraphics[width = 8cm]{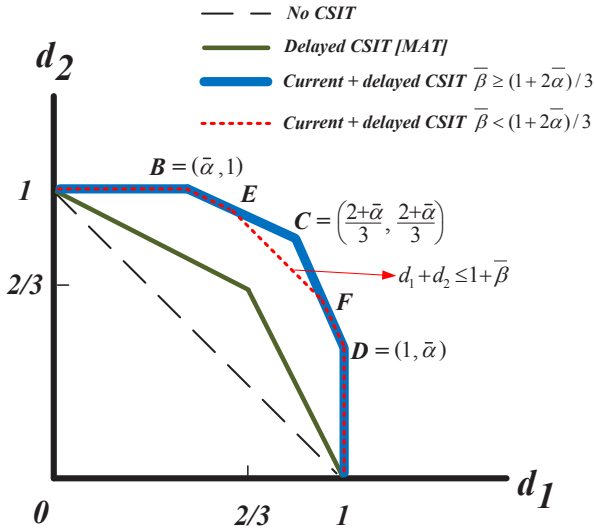}
\caption{DoF region of two-user MISO BC with symmetric feedback, $\bar{\alpha}^{(1)}= \bar{\alpha}^{(2)}=\bar{\alpha}, \bar{\beta}^{(1)}= \bar{\beta}^{(2)}=\bar{\beta}$. The optimal region takes the form of a polygon with corner points $\{(0,0),(0,1),(\bar{\alpha},1),(\frac{2+\bar{\alpha}}{3},\frac{2+\bar{\alpha}}{3}),(1,\bar{\alpha}),(1, 0)\}$ for $\bar{\beta}\geq \frac{1+2\bar{\alpha}}{3}$. For $\bar{\beta}<\frac{1+2\bar{\alpha}}{3}$, the derived region takes the form of a polygon with corner points $\{(0,0),(0,1),(\bar{\alpha},1),(2\bar{\beta}-\bar{\alpha},1+\bar{\alpha}-\bar{\beta}),(1+\bar{\alpha}-\bar{\beta}, 2\bar{\beta}-\bar{\alpha}),(1,\bar{\alpha}),(1, 0)\}.$}
\label{fig:DoFR_SymmCSIT_IDCSIT}
\end{figure}

Figure~\ref{fig:DoFR_SymmCSIT_IDCSIT} depicts the DoF region of the two-user MISO BC in the presence of CSIT feedback with long-term symmetry.

We now quantify the extent to which having symmetric feedback offers a benefit over the asymmetric case where one user accumulates - in the long term - better feedback than the other. Such `symmetry gains' have been recorded in different instances (cf.~\cite{TJSP:12},\cite{MJS:12}).

The following broad comparison focuses on the case of perfect delayed CSIT ($\bar{\beta} = 1$), and contrasts the symmetric feedback case $\bar{\alpha}^{(1)}=\bar{\alpha}^{(2)}$, to the asymmetric case $\bar{\alpha}^{(1)}\neq\bar{\alpha}^{(2)}$. Naturally such comparison is performed under an overall feedback constraint, which - reflecting the spirit of previous works that have identified symmetry gains - is here chosen to be in the form of a fixed sum $\bar{\alpha}^{(1)}+\bar{\alpha}^{(2)}$.
The comparison is in terms of the optimal sum DoF $d_1+d_2$, where again we recall that the users are labeled so that $\bar{\alpha}^{(1)} \geq \bar{\alpha}^{(2)}$.
To clarify, the symmetry gain will be the difference in the sum-DoF performance of two cases; the symmetric case where the two exponent averages are the same and are equal to $\frac{1}{2}(\bar{\alpha}^{(1)}+\bar{\alpha}^{(2)})$, and the asymmetric case where the two distinct exponent averages are $\bar{\alpha}^{(1)}$ and $\bar{\alpha}^{(2)}$.
The proof is direct from Theorem~\ref{thm:genCSIT} and Corollary~\ref{cor:sym}.

\vspace{3pt}
\begin{corollary}[Symmetric vs. asymmetric feedback] \label{cor:Asyvssym}
The symmetry sum-DoF gain is equal to $\frac{1}{6}(2\bar{\alpha}^{(1)} - \bar{\alpha}^{(2)} - 1)^{+}$, i.e., if $2\bar{\alpha}^{(1)} - \bar{\alpha}^{(2)} - 1 > 0$, the symmetric sum-DoF gain is $\frac{2\bar{\alpha}^{(1)} - \bar{\alpha}^{(2)} - 1}{6}>0$, else there is no symmetry gain.
\end{corollary}
\vspace{3pt}

\begin{example}
For example, consider the asymmetric feedback option $\bar{\alpha}^{(1)}=1, \bar{\alpha}^{(2)}=0$
which corresponds to an optimal sum-DoF of $d_1+d_2 = 3/2$ (see Theorem~\ref{thm:genCSIT}, and consider perfect delayed CSIT), and compare this with the symmetric option where both exponent averages are equal to $1/2$. The symmetric option provides a sum-DoF of $d_1+d_2 = 5/3$, and a symmetry gain of $5/3 - 3/2 = 1/6$. As expected, the gain is positive since $2\bar{\alpha}^{(1)} - \bar{\alpha}^{(2)} - 1 =2 - 0 - 1 =1 > 0$.

On the other hand, an asymmetric option $\bar{\alpha}^{(1)}=3/5, \bar{\alpha}^{(2)}=2/5$ corresponds to an optimal sum DoF of $d_1+d_2 = 5/3$, which matches the aforementioned DoF performance of the symmetric option. The symmetry gain here is zero, since $2\bar{\alpha}^{(1)} - \bar{\alpha}^{(2)} - 1 =6/5  - 2/5 - 1 = -1/5< 0$.
\end{example}
\vspace{3pt}

Finally, before concluding our discussion on feedback symmetry/asymmetry, it is worth noting that the asymmetric setting here - where $\bar{\alpha}^{(1)}\neq \bar{\alpha}^{(2)}$ and where $\bar{\beta}^{(1)}$ and $\bar{\beta}^{(2)}$ need not be equal - yields a natural generalization for the asymmetric setting of Maleki et al. in~\cite{MJS:12} which, as we have mentioned, in the presence of abundant delayed CSIT, had an optimal DoF corresponding to DoF corner point $(1,1/2)$ (and a sum-DoF $d_1+d_2 = 3/2$).
The following corollary - which again corresponds to the range of sufficiently good delayed CSIT where $\min \{ \beta^{(1)}, \beta^{(2)}\} \geq \min \{\frac{1+\bar{\alpha}^{(1)}+\bar{\alpha}^{(2)}}{3}, \frac{1+\bar{\alpha}^{(2)}}{2}\}$ - offers a broad generalization of the corresponding result in~\cite{MJS:12}. The proof is direct from the main result.

\vspace{3pt}
\begin{corollary}[Asymmetric and periodic CSIT] \label{cor:MalekiGeneralization}
In the range of sufficiently good delayed CSIT, the optimal DoF region is defined by corner points $B=(\bar{\alpha}^{(2)}, 1)$, $C=(\frac{2+2\bar{\alpha}^{(1)}-\bar{\alpha}^{(2)}}{3}, \frac{2+2\bar{\alpha}^{(2)}-\bar{\alpha}^{(1)}}{3})$ and $D=(1, \bar{\alpha}^{(1)})$ whenever $2\bar{\alpha}^{(1)} - \bar{\alpha}^{(2)}< 1$, else by corner points
$A=(1, \frac{1+\bar{\alpha}^{(2)}}{2})$ and $B$.
\end{corollary}
\vspace{3pt}

As an example we can see that the same DoF corner point $A = (1,1/2)$ - derived in~\cite{MJS:12} under the general principle of perfect delayed CSIT for both users, and perfect current CSIT for the first user - can in fact be achieved with a plethora of options with lesser current and delayed CSIT, such as
\[ \alpha^{(1)}_t = 1/2, \ \alpha^{(2)}_t = 0 , \  \beta^{(1)}_t = \beta^{(2)}_t = 1/2, \ \forall t .\]


\subsubsection{Need for delayed feedback: Imperfect vs. perfect delayed CSIT \label{sec:ImperfectDelayedCSIT}}

We now shift emphasis to explore the fact that imperfect delayed CSIT $(\bar{\beta}<1)$ can - in some cases - be as useful as (asymptotically) perfect delayed CSIT $(\bar{\beta}=1)$, and to provide insight on the overall feedback quality (timely and delayed) that is necessary to achieve a certain DoF performance.

Before proceeding with the result, we briefly motivate our interest in imperfect and sparse \emph{delayed} CSIT.  Towards this we recall that $\bar{\alpha}^{(1)},\bar{\alpha}^{(2)}$ are more representative of the quality (and inevitably of the amount) of \emph{timely} feedback, while $\bar{\beta}^{(1)},\bar{\beta}^{(2)}$ are more representative of the quality of the \emph{entirety} of feedback (timely plus delayed). In this sense, any attempt to limit the total amount and quality of feedback - that is communicated during a certain communication process - must include reducing $\bar{\beta}^{(1)},\bar{\beta}^{(2)}$, rather than just focusing on reducing $\bar{\alpha}^{(1)},\bar{\alpha}^{(2)}$.
For example, even if we removed entirely all current CSIT ($\alpha^{(1)}_t=\alpha^{(2)}_t=0, \forall t$), but insisted on always sending perfect delayed CSIT ($\beta^{(1)}_t=\beta^{(2)}_t=1, \forall t$), we would achieve little towards reducing the total amount of feedback, and we would mainly shift the time-frame of the problem, again irrespective of the drastic reduction in $\bar{\alpha}^{(1)},\bar{\alpha}^{(2)}$.

As we will see though, having reduced $\bar{\alpha}^{(1)},\bar{\alpha}^{(2)}$ can in fact translate to having overall reduced feedback because, interestingly, having reduced $\bar{\alpha}^{(1)},\bar{\alpha}^{(2)}$, can translate - to a certain extent - to needing lesser quality delayed feedback, i.e., can translate to further reductions in $\bar{\beta}^{(1)},\bar{\beta}^{(2)}$. This is quantified in the following, the proof of which is direct, because it simply restates part of what is in the theorem.

\vspace{3pt}
\begin{corollary} [Imperfect vs. perfect delayed CSIT] \label{cor:ImperfectVsPerfectDelayed}
A CSIT process $\{\hat{\bfh}_{t,t'},\hat{\bfg}_{t,t'}\}_{t=1,t'=1}^n$ that offers \beq \label{eq:betaThresholdAsym}\min \{ \bar{\beta}^{(1)}, \bar{\beta}^{(2)}\} \geq \min \{\frac{1+\bar{\alpha}^{(1)}+\bar{\alpha}^{(2)}}{3}, \frac{1+\bar{\alpha}^{(2)}}{2}\}\eeq gives the same DoF as a CSIT process that offers perfect delayed CSIT for each channel realization ($\beta^{(1)}_t=\beta^{(2)}_t=1, \forall t$, i.e., $\bar{\beta}^{(1)}=\bar{\beta}^{(2)}=1$).\\
For the symmetric case, having \beq \label{eq:betaThresholdSym}\bar{\beta} \geq \frac{1+2\bar{\alpha}}{3}\eeq guarantees the same.
\end{corollary}
\vspace{3pt}

It is interesting to observe that the expressions in the above corollary match the amount of delayed CSIT used by schemes in the past, even though such schemes were not designed with the expressed purpose of reducing the amount of delayed CSIT.
For example, the Maddah-Ali and Tse scheme in \cite{MAT:11c} (feedback with $\alpha^{(1)}_t = \alpha^{(2)}_t = 0,\forall t$, over an i.i.d fast-fading channel), while in principle corresponding to abundant delayed CSIT, in fact was based on a precoding design that only needed delayed CSIT only for every third channel realization, corresponding to
\beq \label{eq:MATDelayedBetas}
\beta^{(i)}_t =
\begin{cases}
  1 & \text{if $t=i \ (\!\!\!\!\mod 3)$}\\
  0 &\text{otherwise}
\end{cases}, \ \text{user} \ i=1,2
\eeq
and\footnote{Here when we say $t=i \ (\!\!\!\!\mod 3)$, we refer to the modulo operation, i.e., we mean that $t = 3k+i$ for some integer $k$.} thus corresponding to $\bar\beta^{(i)} = 1/3, \ i=1,2$, which happens to match the above expression in~\eqref{eq:betaThresholdSym} ($\bar{\alpha} = 0$).  This same general expression in~\eqref{eq:betaThresholdSym} additionally tells us that, in the Maddah-Ali and Tse setting, any combination of CSIT quality exponents that allows for $\bar\beta^{(1)} =\bar\beta^{(2)} \geq 1/3$, will allow for the same optimal DoF region in~\cite{MAT:11c}. For example, one such choice would be to use $\beta^{(1)}_t = \beta^{(2)}_t = 1/3,\forall t$.

A similar observation holds for the optimal schemes in~\cite{YKGY:12d,GJ:12o} ($\alpha^{(1)}_t = \alpha^{(2)}_t = \alpha,\forall t$) which again operated in a setting that in principle allowed for unlimited delayed CSIT, but which in fact asked for delayed CSIT only for every third channel realization
\beq \label{eq:ShengDelayedBetas}
\beta^{(i)}_t =
\begin{cases}
  1 & \text{if $t=i \ (\!\!\!\!\mod 3)$}\\
  \alpha &\text{otherwise}
\end{cases}
\eeq
corresponding to $\bar\beta^{(i)} = (1+2\alpha)/3, \ i=1,2$.  This again can be seen as a special instance of the general expression in~\eqref{eq:betaThresholdSym}, which is powerful enough to reveal that any combination of CSIT quality exponents that allows for $\bar\beta^{(1)} =\bar\beta^{(2)} \geq (1+2\bar{\alpha})/3$, will achieve the same optimal DoF in~\cite{YKGY:12d,GJ:12o}. One such choice would be to have $\beta^{(1)}_t = \beta^{(2)}_t = \frac{1+2\alpha}{3},\forall t$.

Along the same lines, the optimal asymmetric scheme in~\cite{MJS:12} which operated under the general principle of perfect delayed CSIT for both users, and perfect current CSIT for the first user, in fact employed a scheme that used lesser feedback. In this scheme, which had duration of two channel uses, the actual required CSIT corresponded to $\alpha_1^{(1)}=\beta_1^{(1)}=\beta_1^{(2)}=1$, and $\alpha_1^{(2)}=\alpha_2^{(1)}=\alpha_2^{(2)}=\beta_2^{(1)}=\beta_2^{(2)}=0$, thus corresponding to $\bar{\alpha}^{(1)}=1/2,\bar{\alpha}^{(2)}=0,\bar{\beta}^{(1)}=1/2,\bar{\beta}^{(2)}=1/2$, which matches the expression in~\eqref{eq:betaThresholdAsym} since $\min\{\bar{\beta}^{(1)},\bar{\beta}^{(2)}\} = \min\{1/2,1/2\} =  \min\{ \frac{1+ \bar{ \alpha}^{(1)} +  \bar{ \alpha}^{(2)}}{3},    \frac{1 +  \min\{ \bar{\alpha}^{(1)},\bar{\alpha}^{(2)}  \}}{2} \}  = \min\{ \frac{1+ 1/2 +  0}{3},   \frac{1+  0}{2} \}  = 1/2$. This same expression in~\eqref{eq:betaThresholdAsym} further reveals other CSIT options that allow for the same optimal DoF.

\subsubsection{Need for predicted CSIT}
We now shift emphasis from delayed CSIT to the other extreme of predicted CSIT.
As we recall, we considered a channel process $\{\bfh_t,\bfg_t\}_t$ and a CSIT process $\{\hat{\bfh}_{t,t'},\hat{\bfg}_{t,t'}\}_{t,t'}$, consisting of estimates $\hat{\bfh}_{t,t'}$ - available at any time $t'$ - of the channel $\bfh_t$ that materializes at any time $t$.  We also advocated that we can safely assume that $\E[||\bfh_t - \hat{\bfh}_{t,t'}||^2]  \leq  \E[||\bfh_t - \hat{\bfh}_{t,t''}||^2] $ (similarly $\E[||\bfg_t - \hat{\bfg}_{t,t'}||^2]  \leq  \E[||\bfg_t - \hat{\bfg}_{t,t''}||^2] $), for any $t'>t''$, simply because one can revert back to past estimates of statistically better quality.
This assumption though does not preclude the possible usefulness of early (predicted) estimates, even if such estimates are generally of lesser quality (statistically) than current estimates (i.e., of lesser quality than estimates that appear during or after the channel materializes).
It is still conceivable that transmission at a certain time $t^{*}$, can benefit from being a function of an estimate $\hat{\bfh}_{t,t'}$ of a future channel $t>t^{*}$, where this estimate became available - naturally by prediction - at any time $t'\leq t^{*}<t$.
The following addresses this, in the range of sufficiently good delayed CSIT where $\min \{ \bar{\beta}^{(1)}, \bar{\beta}^{(2)}\} \geq \min \{\frac{1+\bar{\alpha}^{(1)}+\bar{\alpha}^{(2)}}{3}, \frac{1+\bar{\alpha}^{(2)}}{2}\}$.

\vspace{3pt}
\begin{corollary}[Need for predicted CSIT]\label{cor:Prediction}
In the range of sufficiently good delayed CSIT, transmission need not consider predicted estimates of future channels, to achieve the optimal DoF.
\end{corollary}
\vspace{3pt}

\begin{proof}
The proof is by construction; the designed schemes do not use predicted estimates, while the tight outer bound does not preclude the use of such predicted estimates.
\end{proof}


\section{Periodically evolving CSIT\label{sec:evolvingCSIT}}

We here focus on the block fading setting with a finite coherence period of $T_c$ channel uses, during which the channel remains fixed, and during which a gradual accumulation of feedback provides a progressively increasing CSIT quality, as time progresses across the coherence period (partially delayed current CSIT), or at any time after the end of the coherence period (delayed and potentially obsolete CSIT)\footnote{This definition of current vs. delayed CSIT, originates from \cite{MAT:11c}, and is the standard definition adopted by most existing works on the topic.}.

Such gradual improvement could be sought in FDD (frequency division duplex) settings with limited-capacity feedback links that can be used more than once during the coherence period to progressively refine CSIT, as well as in TDD (time division duplex) settings that use reciprocity-based estimation that progressively improves over time.

In this setting, where the channel remains the same for a finite duration of $T_c$ channel uses, the time index is arranged so that
\[  \bfh_{\ell T_c+1} = \bfh_{\ell T_c+2} = \cdots =\bfh_{(\ell+1) T_c}\]
\[  \bfg_{\ell T_c+1} = \bfg_{\ell T_c+2} = \cdots =\bfg_{(\ell+1) T_c}\]
for a non-negative integer $\ell$. As a result, in the presence of a periodic feedback process which repeats with period $T_c$, we are presented with a periodic sequence of current-CSIT quality exponents
\beq \label{eq:periodicallyEvolvingExponents}
\alpha_{t}^{(i)} = \alpha_{\ell T_c+t}^{(i)}, \forall \ell = 0,1,2,\cdots, \ i=1,2.
\eeq
We focus here - simply for the sake of clarity of exposition - on the symmetric feedback case ($\bar{\alpha}^{(1)}= \bar{\alpha}^{(2)}=:\bar{\alpha}$). In this setting - and after adopting a periodic time index corresponding to having $\ell=0$ (cf.~\eqref{eq:periodicallyEvolvingExponents}) - the time horizon of interest spans $t=1,2,\cdots,T_c$, and the feedback quality is now represented by the $T_c$ current CSIT quality exponents $\{\alpha_t\}_{t=1}^{T_c}$ and by the delayed CSIT exponent $\beta$. Specifically each $\alpha_t$ describes the high SNR precision of the current CSIT estimates at time $t\leq T_c$, whereas $\beta$ captures the precision of the best CSIT estimate received after the channel has elapsed, i.e., after the coherence period of the channel.
In this setting we have that
\beq
\label{eq:exponentIneq}
0\leq \alpha_1\leq \cdots \leq \alpha_{T_c}\leq \beta \leq 1
\eeq
where - since the channel remains fixed during the coherence period - any difference between two consecutive exponents is attributed to feedback that was received during that time slot.

One of the utilities of this setting is that it concisely captures practical timing issues, capturing the effects of feedback that offers an inaccurate representation of the true state of the channel, as well the effects of feedback that can only be used for a small fraction of the communication duration.
Having for example $\alpha_1 = 1$ simply refers to the case of asymptotically perfect and immediately available (full) CSIT, whereas having $\alpha_{T_c} = 0$ simply means that no (or very limited) current feedback is sent during the coherence period of the channel.  Similarly having $\alpha_{\gamma T_c} = 0,$ for some $\gamma\in[0,1]$, simply means that no (or very limited) current feedback is sent during the first $\gamma$ fraction of the coherence period\footnote{Our ignoring here of integer rounding is an abuse of notation that is only done for the sake of clarity of notation, and it carries no real effect on the result.}.
\vspace{3pt}

\begin{example}
Having a periodic feedback process that sends refining feedback, let's say, two times per coherence period, at times $t = \gamma_1 T_c+1,t = \gamma_2 T_c+1$ and never again about that same channel, will result in having
\begin{multline}
\overbrace{0 = \alpha_1 = \cdots = \alpha_{\gamma_1 T_c}}^{\text{Before feedback}} \leq \overbrace{\alpha_{\gamma_1 T_c+1} = \cdots = \alpha_{\gamma_2 T_c}}^{\text{After first feedback}} \\ \leq
\underbrace{\alpha_{\gamma_2 T_c+1} = \cdots = \alpha_{T_c} = \beta}_{\text{After second feedback}}
\end{multline}
whereas if the same feedback system is modified to further add some delayed feedback after the channel elapses, may allow for $\beta > \alpha_{T_c}$.
\end{example}
\vspace{3pt}

One can note that reducing $\alpha_{T_c}$, implies a reduced amount of feedback - about a specific channel - that is sent during the coherence period of that same channel. On the other hand, reducing $\beta$ implies a reduced amount of feedback, during and after the channel's coherence period. Along these lines, reducing $(\beta - \alpha_{T_c})$ implies a reduced amount of feedback, about a specific fading coefficient, that is sent after the coherence period of the channel.

The results here hold directly from the previous results in this work, where directly from \eqref{eq:averages}, we now simply have that
\beq \label{eq:averageAlphaEvolving}
\bar{\alpha}  = \frac{1}{T_c}\sum_{t=1}^{T_c} \alpha_t.\eeq
The following - which is placed here for completeness - holds directly from Corollary~\ref{cor:sym}, for the case of a periodically evolving feedback process over a quasi-static channel.

\vspace{3pt}
\begin{corollary}[Periodically evolving feedback]\label{cor:evolvingSymmetric}
For a periodic feedback process with $\{\alpha_t\}_{t=1}^{T_c}$ and perfect delayed CSIT (received at any time after the end of the coherence period), the optimal DoF region over a block-fading channel is the polygon with corner points \beq \label{eq:DoFSymmetricEvolving}
\{(0,0), (0,1), (\bar{\alpha},1), (\frac{2+\bar{\alpha}}{3},\frac{2+\bar{\alpha}}{3}),(1,\bar{\alpha}),(1, 0)\}.\eeq
This same optimal region can in fact be achieved even with imperfect-quality delayed CSIT, as long as $\beta\geq \frac{1+2\bar{\alpha}}{3}$.
\end{corollary}
\vspace{3pt}

\begin{remark}[Feedback quality vs. quantity]
While all the results here are in terms of feedback \emph{quality} rather than in terms of feedback \emph{quantity}, there are distinct cases where the relationship between the two is well defined.  Such is the case when CSIT estimates are derived using basic - and not necessarily optimal - scalar quantization techniques~\cite{CT:06}.  In such cases, which we mention here simply to offer some insight\footnote{We clarify that this relationship between CSIT quality and feedback quantity, plays no role in the development of the results, and is simply mentioned in the form of comments that offer intuition.  Our focus is on quality exponents, and we make no optimality claim regarding the number of quantization bits.} - and remaining in the high SNR regime - dedicating $\alpha \log P$ quantization bits, per scalar, to quantize $\bfh$ into an estimate $\hat{\bfh}$, allows for a mean squared error~\cite{CT:06} \[\E \|\bfh-\hat{\bfh}\|^2 \doteq P^{-\alpha}. \]
\end{remark}

Drawing from this, and going back to our previous example, let us consider a similar example.
\vspace{3pt}

\begin{example}
Consider a periodic feedback process that sends refining feedback two times per coherence period, by first sending $\alpha' \log P$ bits of feedback per scalar at time $t = \gamma_1 T_c+1$, then by sending extra $\alpha'' \log P$ bits of feedback per scalar at time $t = \gamma_2 T_c+1$, and where it finally sends $(\beta-(\alpha'+\alpha''))\log P$ extra bits of refining feedback per scalar, at some fixed point in time after the end of the coherence period of the channel. This would result in having
\begin{multline}
\overbrace{0 = \alpha_1 = \cdots = \alpha_{\gamma_1 T_c}}^{\text{Before feedback}} \leq \overbrace{\alpha' = \alpha_{\gamma_1 T_c+1} = \cdots = \alpha_{\gamma_2 T_c}}^{\text{After first feedback}} \\ \leq
\underbrace{\alpha'+\alpha'' = \alpha_{\gamma_2 T_c+1} = \cdots = \alpha_{T_c}}_{\text{After second feedback, before $T_c$}} \leq \underbrace{\beta}_{\text{After coherence period}}.
\end{multline}

For instance, if this periodic feedback process sends $\frac{4}{9} \log P$ feedback bits per scalar, at time $t = \frac{1}{3} T_c+1$, and then sends extra $\frac{1}{9} \log P$ bits of feedback at time $t = \frac{2}{3} T_c+1$, it will allow for
\begin{multline}
\overbrace{0 = \alpha_1 = \cdots = \alpha_{\frac{1}{3}T_c}}^{\text{Before feedback}} \leq \overbrace{\frac{4}{9} = \alpha_{\frac{1}{3} T_c+1} = \cdots = \alpha_{\frac{2}{3} T_c}}^{\text{After first feedback}} \\ \leq
\underbrace{\frac{5}{9} = \alpha_{\frac{2}{3} T_c+1} = \cdots = \alpha_{T_c}}_{\text{After second feedback, before $T_c$}}
\end{multline}
which gives $\bar{\alpha} = (0+4/9+5/9)/3 = 1/3$, which in turn gives (Corollary~\ref{cor:evolvingSymmetric}) an optimal DoF region which is defined by the polygon with corner points
\beq
\{(0,0), (0,1), (1/3,1), (7/9,7/9),(1,1/3),(1, 0)\}.\eeq
\end{example}
\vspace{3pt}

Note that in this example, there is no need for extra bits of (delayed) feedback after the end of the coherence period, because the existing amount and timing of feedback bits - again under scalar quantization - guarantees that \[ \beta = \alpha_{T_c} =  5/9= \frac{1+2\bar{\alpha}}{3}= \frac{1+2/3}{3}\] which we have seen (Corollary~\ref{cor:evolvingSymmetric}) to already be as good as perfect delayed feedback ($\beta=1$).
\vspace{3pt}

Placing our focus back on feedback quality, and remaining on the setting of periodically evolving feedback, we proceed with a corollary that offers insight on the question of what CSIT quality and timing, suffice to achieve a certain DoF performance. For ease of exposition, we focus on the hardest-to-achieve DoF point $d_1 = d_2 = d$. The proof is again direct.

\vspace{3pt}
\begin{corollary}[Sufficient feedback for target DoF] \label{cor:HowMuchAndWHenForDOF}
Having $\bar{\alpha} \geq 3d-2$ with $\beta\geq 2d-1$, or having $\bar{\alpha} \geq 3d-2$ with $\alpha_{T_c} \geq 2d-1$ (and no extra delayed feedback), suffices to achieve a symmetric target DoF $d_1 = d_2 = d$.
\end{corollary}
\vspace{3pt}

One can see that having $\bar{\alpha} \geq 3d-2$ with $\alpha_{T_c} \geq 2d-1$ simply means that there is no need to send delayed feedback, i.e., there is no need to send feedback after the end of the coherence period.

Another practical aspect that is addressed here - again in the context of periodically evolving feedback - has to do with feedback delays.  Such delays might cause performance degradation, which might be mitigated if the feedback - albeit with delays - has higher precision.
The following corollary provides some insight on these aspects, by describing the feedback delays that allow a given target symmetric DoF $d$ in the presence of constraints on current and delayed CSIT qualities.
We will be specifically interested in the allowable fractional delay of feedback (cf.~\cite{LH:12})
 \beq \label{eq:gamma} \gamma\defeq \arg\max_{\gamma'} \{\alpha_{\gamma' {T_c}} = 0 \}\eeq
i.e., the fraction $\gamma\leq 1$ for which $\alpha_1 = \cdots = \alpha_{\gamma {T_c}} = 0, \alpha_{\gamma {T_c}+1} > 0$.
We are also interested to see how this allowable delay reduces in the presence of a constraint $\alpha_t\leq \alpha_{\max} \forall t$ on timely feedback, or in the presence of a constraint on $\beta$.

\vspace{3pt}
\begin{corollary}[Allowable feedback delay]\label{cor:delayWithConstraints}
Under a current CSIT quality constraint $\alpha_t\leq \alpha_{\max} \  \forall t$, a symmetric target DoF $d$ can be achieved with any fractional delay
\begin{align*}
\gamma \leq
\begin{cases}
  1- \frac{3d-2}{\alpha_{\max}}        & \quad \text{if \  $d\in [2/3,  (2+\alpha_{\max})/3]$}   \\
  1   &  \quad   \text{if  \ $d\in [0, 2/3]$}
\end{cases}
\end{align*}
while under a constraint $\beta\leq \beta_{\max}$, it can be achieved with any
\begin{align*}
\gamma \leq
\begin{cases}
  1   &     \text{if  \ $d\in [0, \frac{1 + \min\{\beta_{\max}, 1/3\}}{2}]$} \\
    \frac{1}{2}\bigl(\frac{1}{2d-1} -1\bigr)       &  \text{else if \  $d\in [\frac{1 + \min\{\beta_{\max}, 1/3\}}{2}, \frac{1+\beta_{\max}}{2}]$}
\end{cases}
\end{align*}
Finally since $\alpha_{\max}\leq \beta_{\max}\leq 1$, the above reveals that under no specific constraint on CSIT quality, $d$ can be achieved with
\begin{align*}
\gamma \leq
\begin{cases}
  3(1-d)      & \quad \text{if \  $d\in [2/3,  1]$}   \\
  1   &  \quad   \text{if  \ $d\in [0, 2/3]$}.
\end{cases}
\end{align*}
\end{corollary}
\vspace{3pt}

To see the above, we first note that in the first case ($\alpha_t\leq \alpha_{\max}$), $d\in [0, 2/3]$ can be achieved by using perfect but delayed feedback sent at any point in time after $t={T_c}$
$$\underbrace{\alpha_1 = \cdots = \alpha_{ T_c} = 0}_{\text{No feedback}}, \ \underbrace{ \beta = 1}_{\text{delayed CSIT at $t >{T_c}$ }}$$
(cf. \cite{MAT:11c}), while $d\in (2/3,  (2+\alpha_{\max})/3]$ can be achieved by setting $\alpha_1 = \cdots = \alpha_{\gamma {T_c}} = 0, \alpha_{\gamma {T_c}+1} = \cdots = \alpha_{T_c} = \alpha_{\max} $,  $\gamma= 1- \frac{3d-2}{\alpha_{\max}} $,   $\beta \geq 2d-1$ (cf.~Corollary~\ref{cor:sym}).

In the second case ($\beta\leq \beta_{\max}$), if $d\in [0, \frac{1 + \min\{\beta_{\max}, 1/3\}}{2}]$, then $d$ can be achieved by using imperfect and delayed feedback sent at any point in time after $t={T_c}$
$$\underbrace{\alpha_1 = \cdots = \alpha_{ T_c} = 0}_{\text{No feedback}}, \ \underbrace{ \beta = \beta_{\max}}_{\text{delayed CSIT at $t >{T_c}$ }}$$
(cf.~Corollary~\ref{cor:sym}), else if $d\in [\frac{2}{3}, \frac{1+\beta_{\max}}{2}]$ and $\beta_{\max} \geq 1/3$, then $d$ can be achieved by setting  $\alpha_1 = \cdots = \alpha_{\gamma {T_c}} = 0, \alpha_{\gamma {T_c}+1} = \cdots = \alpha_{T_c} = \beta = 2d -1$, $\gamma= \frac{1}{2}\bigl(\frac{1}{2d-1} -1\bigr) $.

Finally, in the unconstrained case, $d\in [0, 2/3]$ can be achieved by setting $\alpha_1 = \cdots  = \alpha_{T_c} =0$ and  $\beta = 1$, while $d\in [2/3,  1]$ can be achieved by using perfect (but partially delayed) feedback sent at $t=\gamma {T_c}+1$
$$\underbrace{\alpha_1 = \cdots = \alpha_{\gamma {T_c}} = 0}_{\text{No feedback}},\underbrace{\alpha_{\gamma {T_c}+1} = \cdots \alpha_{T_c} = \beta = 1}_{\text{Perfect quality CSIT}}.$$

\begin{example}
Consider a symmetric target DoF of $d_1 = d_2 = d=\frac{7}{9}$.
This can be achieved with $\gamma = 3(1-d) = 2/3$ if there is no bound on the quality exponents, and with $\gamma = 1- (3d-2)/\alpha_{\max} = 1/3$ if the feedback link only allows for $\alpha_t \leq \alpha_{\max} = 1/2, \ \forall t$.
If on the other hand, feedback timeliness is easily obtained, we can substantially reduce the amount of CSIT and achieve the same $d=\frac{7}{9}$ with $\alpha_1 = \cdots = \alpha_{T_c} = \bar{\alpha} = 3d-2 = 1/3$ ($\gamma=0,\beta = \frac{1+2\bar{\alpha}}{3} = 2d-1 = 5/9$).
\end{example}

\section{Universal encoding-decoding scheme\label{sec:schemes}}

We proceed to describe the universal scheme that achieves the aforementioned DoF corner points.
The challenge entails designing a scheme of an asymptotically large duration $n$, that utilizes a CSIT process $\{\hat{\bfh}_{t,t'},\hat{\bfg}_{t,t'}\}_{t=1,t'=1}^n$ of quality
defined by the statistics of $\{(\bfh_t-\hat{\bfh}_{t,t'}),(\bfg_t-\hat{\bfg}_{t,t'})\}_{t=1,t'=1}^n$.  This will be achieved by focusing on the corresponding quality-exponent sequences $\{\alpha_{t}^{(1)}\}_{t=1}^n,\{\alpha_{t}^{(2)}\}_{t=1}^n,\{\beta_{t}^{(1)}\}_{t=1}^n,\{\beta_{t}^{(2)}\}_{t=1}^n$, as these were defined in \eqref{eq:alpha1}-\eqref{eq:beta2}.  The optimal DoF region in Theorem~\ref{thm:genCSIT} and the additional corner points in Proposition~\ref{prop:genCSITInner}, will be achieved by properly utilizing different combinations of zero forcing, superposition coding, interference compressing and broadcasting, as well as proper power and rate allocation.
\paragraph{Phase-Markov forward-backward scheme}
Building on the phase-Markov ideas of \cite{CG:79,Due:80} as well as on the ideas of retrospective interference alignment in \cite{MAT:11c} and the ideas of interference quantizing and forwarding in \cite{KYGY:12o}, the current scheme has a forward-backward phase-Markov structure which, in the context of imperfect and delayed CSIT, was first introduced in~\cite{CE:12d,CE:12c} to consist of four main ingredients: block-Markov encoding, spatial precoding, interference quantization and forwarding, and backward decoding.

The scheme asks that the accumulated quantized interference bits of a certain (current) phase, be broadcasted to both users inside the common information symbols of the next phase, while also a certain amount of common information can be transmitted to both users during the current phase, which will then help resolve the accumulated interference of the previous phase.
\vspace{2pt}

As previously suggested, this causal scheme does not require knowledge of future quality exponents, nor does it use predicted CSIT estimates of future channels. The transmitter must know though the long term averages $\bar{\alpha}^{(1)}, \bar{\alpha}^{(2)},\bar{\beta}^{(1)},  \bar{\beta}^{(2)}$, which - as is commonly assumed of long term statistics - can be derived.

By `feeding' this universal scheme with the proper parameters, we can get schemes that are tailored to the different specific settings we have discussed.  We will see such examples later in this section.

We remind the reader that the users are labeled so that $\bar{\alpha}^{(2)}\leq \bar{\alpha}^{(1)}$.
We also remind the reader of the soft assumption that any sufficiently long subsequence $\{\alpha^{(1)}_t\}_{t=\tau}^{\tau+T}$ (resp. $\{\alpha^{(2)}_t\}_{t=\tau}^{\tau+T},\{\beta^{(1)}_t\}_{t=\tau}^{\tau+T},\{\beta^{(2)}_t\}_{t=\tau}^{\tau+T}$) is assumed to have an average that converges to the long term average $\bar{\alpha}^{(1)}$ (resp. $\bar{\alpha}^{(2)},\bar{\beta}^{(1)},\bar{\beta}^{(2)}$), for a finite $T$ that can be sufficiently large to allow for this convergence.
We briefly note that, as we will see later, in periodic settings such as those described in Section~\ref{sec:evolvingCSIT}, $T$ need not be large.

We proceed to describe in Section~\ref{sec:scheme_en} the encoding part, and in Section~\ref{sec:scheme_de} the decoding part.  In Section~\ref{sec:scheme_DoF} we show how the scheme achieves the different DoF corner points of interest.  Finally in Section~\ref{sec:scheme_exam} we provide example instances of our general scheme, for specific cases of particular interest.

For notational convenience, we will use
\[\hat{\hv}_t \defeq \hat{\hv}_{t,t},  \quad \hat{\gv}_t \defeq \hat{\gv}_{t,t}\]
\[\check{\hv}_t \defeq \hat{\hv}_{t,t+\eta},  \quad \check{\gv}_t \defeq \hat{\gv}_{t,t+\eta }\]
to denote the current and delayed estimates of $\hv_t$ and $\gv_t$, respectively\footnote{Recall that $\eta$ is a sufficiently large but finite integer, corresponding to the maximum delay allowed for waiting for delayed CSIT.}, with corresponding estimation errors being
\begin{equation}
\label{eq:MMSEc}
  \tilde{\hv}_t  \defeq \hv_t - \hat{\hv}_t, \quad     \tilde{\gv}_t  \defeq \gv_t - \hat{\gv}_t
\end{equation}
\begin{equation}
\label{eq:MMSEd}
  \ddot{\hv}_t \defeq \hv_t - \check{\hv}_t, \quad     \ddot{\gv}_t  \defeq \gv_t - \check{\gv}_t.
\end{equation}

\subsection{Scheme $\Xc$: encoding \label{sec:scheme_en}}
Scheme $\Xc$ is designed to have $S$ phases, where each phase has a duration of $T$ channel uses, and where $T$ is finite but - unless stated otherwise - sufficiently large.
Specifically each phase~$s$ ($s = 1,2,\cdots,S$) will take place over all time slots $t$ belonging to the set
\begin{align}
\Bc_s = \{\Bc_{s,\ell}  \defeq (s\!-\!1)2T +\ell  \}_{\ell=1}^{T}, \quad   s=1,\cdots,S. \label{eq:timephase}
\end{align}
As stated, $T$ is sufficiently large so that
\begin{align} \label{eq:average}
\frac{1}{T}\sum_{t\in \Bc_s} \alpha^{(i)}_t \!\rightarrow\!  \bar{\alpha}^{(i)} , \  \frac{1}{T}\sum_{t\in \Bc_s}  \beta^{(i)}_t \!\rightarrow\!  \bar{\beta}^{(i)},  \ s=1,\cdots,S
\end{align}
$i=1,2$.
The above allocation in~\eqref{eq:timephase} guarantees that there are $T$ channel uses in between any two neighboring phases.  Having $T$ being sufficiently large allows for the delayed CSIT corresponding to the channels appearing during phase~$s$, to be available before the beginning of the phase that we label as phase $s+1$.
This implies that $T>\eta$ (cf.~\eqref{eq:beta1},\eqref{eq:beta2}), although this assumption can be readily removed\footnote{The assumption can be removed because we can, instead of splitting time into two interleaved halves and identifying each half to a message, to instead split time into more parts, each corresponding to a different message.  For a sufficiently large number of parts, this would allow for the removal of the assumption that $T\geq \eta$, and the only assumption that would remain would be that $T$ is large enough so that~\eqref{eq:average} is satisfied. In periodic settings, such $T$ can be small.}.
Naturally there is no silent time, and over the remaining channel uses $$t \in { \{(2s-1)T+\ell\}_{\ell=1,s=1}^{\ell=T,s=S} }$$ we simply repeat scheme $\Xc$ with a different message. With $n$ being generally infinite, $S$ is also infinite (except for specific instances, some of which are highlighted in Section~\ref{sec:scheme_exam}).

\begin{figure*}[t]
\centering
\includegraphics[width=13cm]{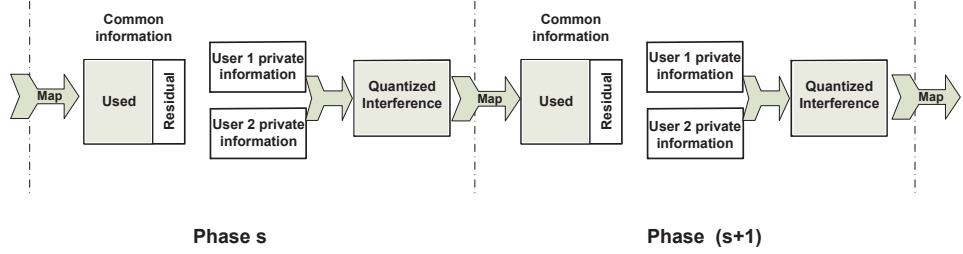}
\caption{Illustration of coding across phases.}
\label{fig:Codingblock}
\end{figure*}

\begin{figure}
\centering
\includegraphics[width=9cm]{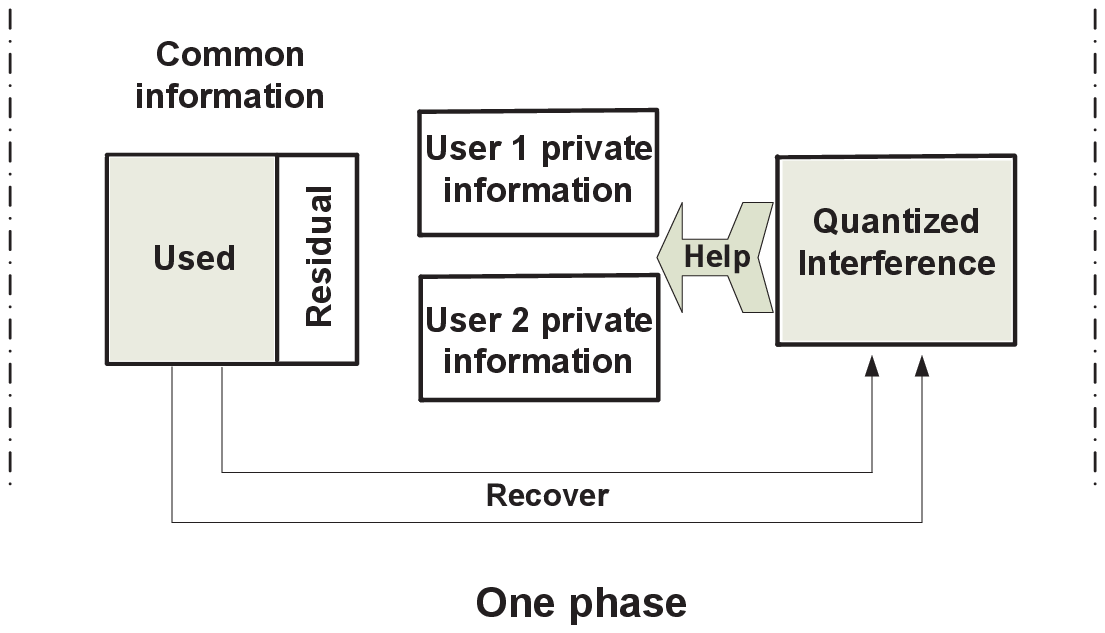}
\caption{Illustration of coding over a single phase.}
\label{fig:Codingblock1phase}
\end{figure}

We proceed to give the general description that holds for all phases $s=1,2,\cdots,S-1$, except for the last phase $S$, which we describe separately afterwards.
A brief corresponding illustration can be found in Figure~\ref{fig:Codingblock} and Figure~\ref{fig:Codingblock1phase}.

\subsubsection{Phase~$s$,  for $s=1,2,\cdots,S-1$}
We proceed to describe the way the scheme, in each phase $s\in[1,S-1]$, combines zero forcing and superposition coding, power and rate allocation, and interference compressing and broadcasting, in order to transmit private information, using currently available CSIT estimates to reduce interference, and using delayed CSIT estimates to alleviate the effect of past interference.

\paragraph{Zero forcing and superposition coding} During phase~$s$, $t\in \Bc_s$, the transmitter sends
\begin{align} \label{eq:TxX1Ph1}
\xv_{t} = \wv_t c_t+\hat{\gv}^{\bot}_t a_t + \hat{\hv}_t a^{'}_t + \hat{\hv}^{\bot}_t b_t + \hat{\gv}_t b^{'}_t \end{align}
where $a_t, a^{'}_t$ are the symbols meant for user 1, $b_t, b^{'}_t$ for user 2, where $c_t$ is a common symbol, where $\ev^{\bot}$ denotes a unit-norm vector orthogonal to $\ev$, and where $\wv_t$ is a predetermined randomly-generated vector known by all the nodes.

\paragraph{Power and rate allocation policy} In describing the power and rates of the symbols in~\eqref{eq:TxX1Ph1}, we use the notation
\beq \label{eq:NotationPower}
P^{(x)}_t \defeq \E |x_t|^2
\eeq to denote the power of $x_t$ corresponding to time-slot~$t$, and we use $r^{(x)}_t$ to denote the prelog factor of the number of bits
$r^{(x)}_t\log P - o(\log P)$ carried by symbol $x_t$ at time $t$.

When in phase $s$, during time-slot $t$, the powers and (normalized) rates are set as
\begin{equation}\label{eq:RPowerX1Ph1}
\begin{array}{cc}
P^{(c)}_t \doteq  P , &  \\
P^{(a)}_t \doteq  P^{\delta_t^{(2)}}, & r^{(a)}_t  = \delta_t^{(2)} \\
P^{(b)}_t \doteq  P^{\delta_t^{(1)}}, & r^{(b)}_t = \delta_t^{(1)} \\
P^{(a')}_t \doteq P^{\delta_t^{(2)}-\alpha^{(2)}_t}, &  r^{(a')}_t =(\delta_t^{(2)}-\alpha^{(2)}_t)^{+} \\
P^{(b')}_t \doteq P^{\delta_t^{(1)}-\alpha^{(1)}_t}, &  r^{(b')}_t =(\delta_t^{(1)}-\alpha^{(1)}_t)^{+} \end{array}
\end{equation}
where $(\bullet)^{+}\defeq\max\{\bullet, 0\}$.

We design the scheme so that the entirety of common information symbols
$\{c_{\Bc_{s,t} }\}_{t=1}^T$, carry
\begin{align}
T(1- \bar{\delta})\log P -o(\log P) \label{eq:ratejointc}
\end{align}
bits, and design the power parameters $\{\delta_t^{(1)}, \delta_t^{(2)}\}_{t\in\Bc_s}$ to satisfy
\begin{align}
\beta_t^{(i)} & \geq \delta_t^{(i)}  \quad  i=1,2,  \ t\in\Bc_s   \label{eq:powerde}\\
\frac{1}{T}\sum_{t\in \Bc_s}\delta_t^{(1)}  & =\frac{1}{T}\sum_{t\in \Bc_s}\delta_t^{(2)} = \bar{\delta}  \label{eq:powerde1} \\
\frac{1}{T}\sum_{t\in \Bc_s} (\delta_t^{(i)}- \alpha_t^{(i)})^{+}  & =(  \bar{\delta} - \bar{\alpha}^{(i)})^{+} \quad i=1,2, \label{eq:powerde2}
\end{align}
for some $\bar{\delta}$ that will be bounded by
\begin{align} \label{eq:bardelta}
\bar{\delta} \leq	\min \{ \bar{\beta}^{(1)},  \bar{\beta}^{(2)}, \frac{ 1 + \bar{\alpha}^{(1)} +  \bar{\alpha}^{(2)}}{3}, \frac{ 1 +\bar{\alpha}^{(2)}}{2}  \}.
\end{align}
There indeed exist solutions $\{\delta_t^{(1)}, \delta_t^{(2)}\}_{t\in\Bc_s}$ that satisfy the above, and an explicit solution is shown in Appendix~\ref{sec:solutionpower}.
Our solution for power and rate allocation allows that, at time $t$, the transmitter needs only acquire knowledge of $\{\alpha_t^{(1)}, \beta_t^{(1)}; \alpha_t^{(2)}, \beta_t^{(2)}\}$, in addition to the derived long-term averages $\bar{\alpha}^{(1)}, \bar{\alpha}^{(2)},\bar{\beta}^{(1)},  \bar{\beta}^{(2)}$.
This nature of the derived solutions is crucial for handling asymmetry ($\alpha^{(1)}_t \neq \alpha^{(2)}_t$, $\beta^{(1)}_t \neq \beta^{(2)}_t$).

After transmission, the received signals take the form
\begin{align}
  y^{(1)}_t&= \underbrace{\hv^\T_t \wv_t c_t}_{P}+\underbrace{\hv^\T_t \hat{\gv}^{\bot}_t a_t}_{P^{\delta_t^{(2)}}} +\underbrace{\hv^\T_t \hat{\hv}_t a^{'}_t}_{P^{\delta_t^{(2)}-\alpha^{(2)}_t}} +\underbrace{z^{(1)}_t}_{P^0} \nonumber\\
  & +\overbrace{\underbrace{\check{\hv}^\T_t ( \hat{\hv}^{\bot}_t b_t+ \hat{\gv}_t b^{'}_t)}_{P^{\delta_t^{(1)}-\alpha^{(1)}_t}}}^{\check{\iota}^{(1)}_t} +  \overbrace{\underbrace{\ddot{\hv}^\T_t ( \hat{\hv}^{\bot}_t b_t+ \hat{\gv}_t b^{'}_t)}_{P^{\delta_t^{(1)}-\beta^{(1)}_t}\leq P^{0}}}^{\iota^{(1)}_t-\check{\iota}^{(1)}_t} \label{eq:sch1y1}\\
  y^{(2)}_t	&= \underbrace{\gv^\T_t \wv_t c_t}_{P}+\underbrace{\gv^\T_t \hat{\hv}^{\bot}_t b_t}_{P^{\delta_t^{(1)}}}+\underbrace{\gv^\T_t \hat{\gv}_t b^{'}_t}_{P^{\delta_t^{(1)}-\alpha^{(1)}_t}}+\underbrace{z^{(2)}_t}_{P^0}\nonumber\\
 &  +\overbrace{\underbrace{\check{\gv}^\T_t ( \hat{\gv}^{\bot}_t a_t+ \hat{\hv}_t a^{'}_t)}_{P^{\delta_t^{(2)}-\alpha^{(2)}_t}}}^{\check{\iota}^{(2)}_t} + \overbrace{\underbrace{\ddot{\gv}^\T_t ( \hat{\gv}^{\bot}_t a_t+ \hat{\hv}_t a^{'}_t)}_{ P^{\delta_t^{(2)}-\beta^{(2)}_t}\leq P^{0}}}^{\iota^{(2)}_t-\check{\iota}^{(2)}_t} \label{eq:sch1y2}
\end{align}
where
\beq \label{eq:c2barsg}
\iota^{(1)}_t  \defeq \hv^\T_t(\hat{\hv}^{\bot}_t b_t + \hat{\gv}_t b^{'}_t), \
\iota^{(2)}_t  \defeq \gv^\T_t (\hat{\gv}^{\bot}_t a_t + \hat{\hv}_t a^{'}_t)
\eeq
denote the interference at user~1 and user~2 respectively, and where
\beq \label{eq:cbarg}
\check{\iota}^{(1)}_t  \defeq \check{\hv}^\T_{t}(\hat{\hv}^{\bot}_t b_t+\hat{\gv}_t b^{'}_t) , \
\check{\iota}^{(2)}_t  \defeq \check{\gv}^\T_{t} (\hat{\gv}^{\bot}_t a_t + \hat{\hv}_t a^{'}_t)
\eeq
denote the transmitter's delayed estimates of the scalar interference terms $\iota^{(1)}_t,\iota^{(2)}_t$.
In the above - where under each term we noted the order of the summand's average power - we considered that
\begin{align}  \label{eq:barcPower1}
\E|\check{\iota}^{(1)}_t|^2\!&=\! \E|\check{\hv}^\T_t \hat{\hv}^{\bot}_t b_t|^2+ \E|\check{\hv}^\T_t \hat{\gv}_t b^{'}_t|^2\nonumber\\
&=\! \E|( \hat{\hv}^\T_t+ \tilde{\hv}^\T_t -\ddot{\hv}^\T_t) \hat{\hv}^{\bot}_t b_t|^2+ \E|\check{\hv}^\T_t \hat{\gv}_t b^{'}_t|^2\nonumber\\
&= \! \E|(\tilde{\hv}^\T_t\!-\!\ddot{\hv}^\T_t )\hat{\hv}^{\bot}_t b_t|^2\!+\! \E|\check{\hv}^\T_t \hat{\gv}_t b^{'}_t|^2 \nonumber\\
& \doteq\!  P^{\delta_t^{(1)}-\alpha^{(1)}_t}\nonumber\\
\E|\check{\iota}^{(2)}_t|^2\!&= \! \E|(\tilde{\gv}^\T_t\!-\!\ddot{\gv}^\T_t )\hat{\gv}^{\bot}_t a_t|^2\!+\! \E|\check{\gv}^\T_t \hat{\hv}_t a^{'}_t|^2 \nonumber\\
& \doteq\!  P^{\delta_t^{(2)}-\alpha^{(2)}_t}.
\end{align}

\paragraph{Quantizing and broadcasting the accumulated interference}  After the end of phase~$s$ and before the beginning of the next phase - which starts $T$ channel uses after the end of phase $s$, i.e., after the accumulation of all delayed CSIT - the transmitter reconstructs $\check{\iota}^{(1)}_t, \check{\iota}^{(2)}_t, t\in \Bc_s$ using its knowledge of delayed CSIT, and quantizes these into
\beq\label{eq:quntisch1}
  \bar{\check{\iota}}^{(1)}_t = \check{\iota}^{(1)}_t -\tilde{\iota}^{(1)}_t, \quad
	\bar{\check{\iota}}^{(2)}_t = \check{\iota}^{(2)}_t - \tilde{\iota}^{(2)}_t
\eeq
with $(\delta_t^{(1)}-\alpha^{(1)}_t)^{+}\log P$ and $(\delta_t^{(2)}-\alpha^{(2)}_t)^{+}\log P$ quantization bits respectively, allowing for bounded power of quantization noise $\tilde{\iota}^{(1)}_t, \tilde{\iota}^{(2)}_t$, i.e, allowing for $$\E|\tilde{\iota}^{(2)}_t|^2 \doteq \E|\tilde{\iota}^{(1)}_t|^2 \doteq 1$$ since $\E|\check{\iota}^{(2)}_t|^2 \doteq P^{\delta^{(2)}_t-\alpha^{(2)}_t}, \ \E|\check{\iota}^{(1)}_t|^2 \doteq P^{\delta^{(1)}_t-\alpha^{(1)}_t}$ (cf.~\cite{CT:06}).
Then the transmitter evenly splits the
\beq\label{eq:bitquan}
\sum_{t\in \Bc_s} \left( (\delta_t^{(1)}-\alpha^{(1)}_t)^{+} + (\delta_t^{(2)}-\alpha^{(2)}_t)^{+} \right)\log P
\eeq
quantization bits into the common symbols $\{c_t\}_{t\in\Bc_{s+1}}$ that will be transmitted during the next phase (phase~$s+1$), conveying these quantization bits together with other new information bits for the users.

This transmission of $\{c_t\}_{t\in\Bc_{s+1}}$ in the next phase, will help each of the users cancel the dominant part of the interference, and it will also serve as an extra observation (see~\eqref{eq:firstMIMOx1} later on) that allows for decoding of all private information of that same user.
Table~\ref{tab:bitsummary} summarizes the number of bits carried by private symbols, common symbols, and by the quantized interference, for phase~$s$, $s=1,2,\cdots,S-1$.

\begin{table}
\caption{Bits carried by private symbols, common symbols, and by the quantized interference, for phase~$s$, $s=1,2,\cdots,S-1$.}
\begin{center}
\begin{tabular}{|c|c|}
  \hline
                & Total bits ($\times \log P$)\\
   \hline
   Private symbols for user~1    & $T(\bar{\delta}+(\bar{\delta}- \bar{\alpha}^{(2)})^{+}  )$ \\
   \hline
   Private symbols for user~2    & $T(\bar{\delta}+(\bar{\delta}- \bar{\alpha}^{(1)})^{+}  )$ \\
   \hline
   Common symbols        &$T(1-\bar{\delta})$  \\
    \hline
  Quantized interference       &$T((\bar{\delta}-\bar{\alpha}^{(1)})^{+} + (\bar{\delta}-\bar{\alpha}^{(2)})^{+})$  \\
    \hline
\end{tabular}
\end{center}
\label{tab:bitsummary}
\end{table}
We now proceed with the description of encoding over the last phase~$S$.

\subsubsection{Phase~$S$}

The last phase, in addition to communicating new private symbols, conveys the remaining accumulated interference from the previous phase, and does so in a manner that allows for termination at the end of this phase.

During this last phase, the transmitter sends
\begin{equation}\label{eq:TxX1PhS}\xv_t =\wv_t c_t+\hat{\gv}^{\bot}_t a_t +\hat{\hv}^{\bot}_t b_t\end{equation}
$t\in \Bc_S$, with power and rates set as
\begin{equation}\label{eq:RPowerX1PhS}
\begin{array}{lll}
P^{(c)}_t \doteq P, &  P^{(a)}_t \doteq  P^{\alpha^{(2)}_t}, &  P^{(b)}_t \doteq  P^{\delta^{(1)}_t} \\
&  r^{(a)}_t = \alpha^{(2)}_t ,& r^{(b)}_t = \delta^{(1)}_t.
\end{array} \end{equation}
With the entirety of common information symbols $ \{c_{\Bc_{S,\ell} }\}_{\ell=1}^T$ now carrying\footnote{We remind the reader of the definition of $\Bc_{s,\ell}$ (cf.~\eqref{eq:timephase}) which denotes the $\ell$th element of set $\Bc_s$ consisting of all time indexes of phase $s$.  For example, saying that $t = \Bc_{1,\ell}$ simply means that $t=\ell$.}
\begin{align} \label{eq:ratecomS}
T(1- \bar{\alpha}^{(2)})\log P -o(\log P)
\end{align}
bits, the power parameters $\{\delta_t^{(1)}\}_{t\in \Bc_S}$ are designed such that
\begin{align}
&\alpha_t^{(1)}  \geq \delta_t^{(1)}  \quad  \forall  t  \label{eq:powerde21}\\
&\frac{1}{T}\sum_{t\in \Bc_S}\delta_t^{(1)} = \bar{\alpha}^{(2)}.  \label{eq:powerde22}
\end{align}
The solution to the above problem is similar to that in \eqref{eq:powerde},\eqref{eq:powerde1},\eqref{eq:powerde2}.

\begin{figure}
\centering
\includegraphics[width=9cm]{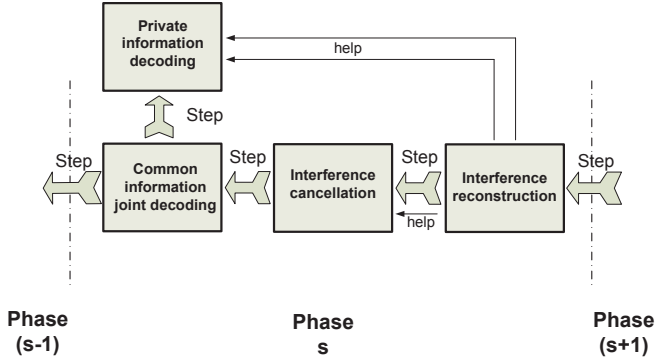}
\caption{Illustration of decoding steps.}
\label{fig:Dedingblock}
\end{figure}

This concludes the part of encoding.  After transmission, the received signals are then of the form
\begin{align}
  y^{(1)}_t	&= \underbrace{\hv^\T_t \wv_t c_t}_{P} +\underbrace{\hv^\T_t \hat{\gv}^{\bot}_t a_t}_{P^{\alpha^{(2)}_t}}  +\underbrace{\tilde{\hv}^\T_t \hat{\hv}^{\bot}_t b_t}_{\leq P^{0}}+\underbrace{z^{(1)}_t}_{P^0} \label{eq:X1PhSy1} \\
  y^{(2)}_t&= \underbrace{\gv^\T_t \wv_t c_t}_{P} +\underbrace{\tilde{\gv}^\T_t\hat{\gv}^{\bot}_t a_t}_{P^{0}}  +\underbrace{\gv^\T_t \hat{\hv}^{\bot}_t b_t}_{P^{\delta^{(1)}_t}}+\underbrace{z^{(2)}_t}_{P^0}. \label{eq:X1PhSy2}
\end{align}
We now move to describe decoding at both receivers, where this decoding part has a phase Markov structure (see Figure~\ref{fig:Dedingblock}), similar to the encoding part.

\subsection{Scheme $\Xc$: decoding \label{sec:scheme_de}}

As it may be apparent (more details will be shown in Section~\ref{sec:scheme_DoF}), the power and rate allocation in \eqref{eq:powerde},\eqref{eq:powerde1},\eqref{eq:powerde2} guarantees that the quantized interference accumulated during phase $s$ ($s=1,\cdots,S-1$) has fewer bits than the load of the common symbols transmitted during the next phase (cf.~\eqref{eq:bitquan}). Consequently decoding of the common symbols during a certain phase, helps recover the interference accumulated during the previous phase. As a result, decoding moves backwards, from the last to the first phase.

\subsubsection{Phase~$S$}

At the end of phase~$S$, we consider \emph{joint} decoding of all common symbols $[c_{\Bc_{S,1}},  \cdots, c_{\Bc_{S,T}}]^\T $. Specifically user~$i$, $i=1,2$, decodes the corresponding common-information vector using its received signal vector $[y^{(i)}_{\Bc_{S,1}},  y^{(i)}_{\Bc_{S,2}}, \cdots, y^{(i)}_{\Bc_{S,T}}]^\T $, and does so by treating the other signals as noise.
We now note that the accumulated mutual information satisfies
\begin{align} \label{eq:sch1ComI}
I([c_{\Bc_{S,1}}, & \cdots, c_{\Bc_{S,T}}]^\T ; [y^{(1)}_{\Bc_{S,1}},  \cdots, y^{(1)}_{\Bc_{S,T}}]^\T)\nonumber\\
&=\log \prod_{t\in \Bc_S} P^{1-\alpha^{(2)}_t} -o(\log P)\nonumber\\
&=T(1-\bar{\alpha}^{(2)})\log P -o(\log P)\nonumber\\
I([c_{\Bc_{S,1}},  &\cdots, c_{\Bc_{S,T}}]^\T ; [y^{(2)}_{\Bc_{S,1}},  \cdots, y^{(2)}_{\Bc_{S,T}}]^\T)\nonumber\\
&=\log \prod_{t\in \Bc_S} P^{1-\delta^{(1)}_t} -o(\log P)\nonumber\\
&=T(1-\bar{\alpha}^{(2)})\log P -o(\log P) \end{align}
 (cf.~\eqref{eq:powerde21},\eqref{eq:powerde22}), to conclude that both users can reliably decode all
\begin{align} \label{eq:sch1Combits}
T(1-  \bar{\alpha}^{(2)}) \log P -o(\log P)
\end{align}
bits in the common information vector $[c_{\Bc_{S,1}},  \cdots, c_{\Bc_{S,T}}]^\T $.
This is proved in Lemma~\ref{lem:latticecode} in the appendix of Section~\ref{sec:DetailsX2}, which in fact guarantees that both users will be able to decode the amount of feedback bits described in \eqref{eq:sch1Combits}, even for finite and small $T$. This is done to ensure the validity of the schemes also for finite $T$, and is achieved by employing specific lattice codes that have good properties in the finite-duration high-SNR regime. The details for this step can be found in the aforementioned appendix.

After decoding $[c_{\Bc_{S,1}},  \cdots, c_{\Bc_{S,T}}]^\T $, user~1 removes $\hv^\T_t \wv_t c_t$ from the received signal in~\eqref{eq:X1PhSy1}, to decode $a_t$.  Similarly user~2 removes $\gv^\T_t \wv_t c_t$ from its received signal in~\eqref{eq:X1PhSy2}, to decode $b_t$.

Now we go back one phase and utilize knowledge of $\{c_t\}_{t\in \Bc_{S}}$, to decode the corresponding symbols.

\subsubsection{Phase~$s$, \ $s= S-1, S-2, \cdots, 1$}

We here describe, for phase~$s$, the actions of interference reconstruction, interference cancelation, joint decoding of common information symbols, and decoding of private information symbols, in the order they happen.

\paragraph{Interference reconstruction}
In this phase (phase~$s$), each user employs knowledge of $\{c_t\}_{t\in \Bc_{s+1}}$ from phase~$s+1$, to reconstruct the delayed estimates of all the interference accumulated in phase~$s$, i.e., to reconstruct $\{\bar{\check{\iota}}^{(2)}_t,\bar{\check{\iota}}^{(1)}_t\}_{t\in \Bc_s}$.

\paragraph{Interference cancelation}
Now with knowledge of $\{\bar{\check{\iota}}^{(2)}_t,\bar{\check{\iota}}^{(1)}_t\}_{t\in \Bc_s}$, each user can remove - up to noise level - all
the interference $\iota^{(i)}_t, \ t\in \Bc_s$, by subtracting the delayed interference estimates $\bar{\check{\iota}}^{(i)}_t$ from $y^{(i)}_t$.

\paragraph{Joint decoding of common information symbols}
At this point, user~$i$ decodes the common information vector $\cv_s \defeq [c_{\Bc_{s,1}},  \cdots, c_{\Bc_{s,T}}]^\T $ from its (modified) received signal vector
$ [y^{(i)}_{\Bc_{s,1}} - \bar{\check{\iota}}^{(i)}_{\Bc_{s,1}},  \cdots, y^{(i)}_{\Bc_{s,T}} - \bar{\check{\iota}}^{(i)}_{\Bc_{s,T}}]^\T$
by treating the other signals as noise.  The accumulated mutual information then satisfies
\begin{align} \label{eq:sch1ComIs}
&I(\cv_s ;[y^{(1)}_{\Bc_{s,1}} - \bar{\check{\iota}}^{(1)}_{\Bc_{s,1}},  \cdots, y^{(1)}_{\Bc_{s,T}} - \bar{\check{\iota}}^{(1)}_{\Bc_{s,T}}]^\T)\nonumber\\
&=\log \prod_{t\in \Bc_s} P^{1-\delta^{(2)}_t} -o(\log P)=T(1-\bar{\delta})\log P -o(\log P)\nonumber\\
&I(\cv_s ;[y^{(2)}_{\Bc_{s,1}} - \bar{\check{\iota}}^{(2)}_{\Bc_{s,1}},  \cdots, y^{(2)}_{\Bc_{s,T}} - \bar{\check{\iota}}^{(2)}_{\Bc_{s,T}}]^\T)\nonumber\\
&=\log \prod_{t\in \Bc_s} P^{1-\delta^{(1)}_t} -o(\log P)=T(1-\bar{\delta})\log P -o(\log P) \end{align}
 (cf.~\eqref{eq:powerde}-\eqref{eq:sch1y2}), and we conclude that both users can reliably decode all \begin{align} \label{eq:sch1Combitss}
T(1- \bar{\delta} ) \log P -o(\log P)
\end{align}
bits of the common information vector $\cv_s$.  The details for this step, can again be found in the appendix of Section~\ref{sec:DetailsX2}.

After decoding $\cv_s $, user~1 removes $\hv^\T_t \wv_t c_t$ from $y^{(1)}_t - \bar{\check{\iota}}^{(1)}_t$, while user~2 removes $\gv^\T_t \wv_t c_t$ from $y^{(2)}_t - \bar{\check{\iota}}^{(1)}_t$, $t\in \Bc_s$.

\paragraph{Decoding of private information symbols}

After removing the interference, and decoding and subtracting out the common symbols, each user now decodes its private information symbols of phase~$s$.
Using knowledge of $\{\bar{\check{\iota}}^{(2)}_t,\bar{\check{\iota}}^{(1)}_t\}_{t\in \Bc_s}$, user~1 will use the estimate $\bar{\check{\iota}}^{(2)}_t$ (of $\check{\iota}^{(2)}_t$) as an extra observation which, together with the observation $y^{(1)}_t-\hv^\T_t \wv_t c_t - \bar{\check{\iota}}^{(1)}_t$, will allow for decoding of both $a_t$ and $a^{'}_t$, $t\in \Bc_s$.
Specifically user~1, at each instance $t$, can `see' a $2\times 2$ MIMO channel of the form
\begin{align} \label{eq:firstMIMOx1}
\begin{bmatrix} y^{(1)}_t-\hv^\T_t\wv_t c_t-\bar{\check{\iota}}^{(1)}_t
          \\ \bar{\check{\iota}}^{(2)}_t \end{bmatrix}   = \begin{bmatrix}  \hv^\T_t \\ \check{\gv}^\T_t \end{bmatrix} \begin{bmatrix}  \hat{\gv}^{\bot}_t  \    \hat{\hv}_t \end{bmatrix} \begin{bmatrix} a_t \\  a^{'}_t \end{bmatrix}
 + {\begin{bmatrix}  \tilde{z}^{(1)}_t\\
              -\tilde{\iota}^{(2)}_t \end{bmatrix}}
\end{align}
where \[\tilde{z}^{(1)}_t= \ddot{\hv}^\T_t (  \hat{\hv}^{\bot}_t  b_t + \hat{\gv}_t b^{'}_t)+ z^{(1)}_t + \tilde{\iota}^{(1)}_t.\]
The fact that $\E|\tilde{z}^{(1)}_t|^2 \doteq 1$, allows for decoding of $a_t$ and $a^{'}_t$, corresponding to the aforementioned rates $r^{(a)}_t = \delta_t^{(2)}, \  r^{(a')}_t = (\delta_t^{(2)}-\alpha^{(2)}_t)^{+}$, $t\in \Bc_s$.
Similar actions are taken by user~2, allowing for decoding of $b_t$ and $b^{'}_t$, again with $r^{(b)}_t = \delta_t^{(1)}, \  r^{(a')}_t = (\delta_t^{(1)}-\alpha^{(1)}_t)^{+}$, $t\in \Bc_s$.

At this point, each user has decoded all the information symbols (common and private) corresponding to phase~$s$, goes back one phase (to phase~$s-1$) to utilize its knowledge of $\{c_t\}_{t\in \Bc_s}$, and decodes the common and private symbols of that phase. The whole decoding effort naturally terminates after decoding of the symbols in the first phase.

\subsection{Scheme $\Xc$: Calculating the achieved DoF\label{sec:scheme_DoF}}
In the following DoF calculation we will consider two separate cases.  Case~1 will correspond to
\begin{align} \label{eq:case1}
2\bar{\alpha}^{(1)}- \bar{\alpha}^{(2)} < 1
\end{align}
which in turn implies that $ \bar{\alpha}^{(1)} \leq \frac{ 1 + \bar{\alpha}^{(1)} +  \bar{\alpha}^{(2)}}{3} \leq \frac{1+  \bar{\alpha}^{(2)}}{2}$, while case~2 will correspond to
\begin{align} \label{eq:case2}
2\bar{\alpha}^{(1)}- \bar{\alpha}^{(2)} \geq 1
\end{align}
which in turn implies that $ \bar{\alpha}^{(1)} \geq \frac{ 1 + \bar{\alpha}^{(1)} +  \bar{\alpha}^{(2)}}{3} \geq \frac{1+  \bar{\alpha}^{(2)}}{2} $.
We recall that the users are labeled so that $\bar{\alpha}^{(1)} \geq \bar{\alpha}^{(2)}$.

\subsubsection{Generic DoF point}
To calibrate the DoF performance, we first note that for any fixed $ \bar{\delta}  \leq	\min \{ \bar{\beta}^{(1)},  \bar{\beta}^{(2)}, \frac{ 1 + \bar{\alpha}^{(1)} +  \bar{\alpha}^{(2)}}{3}, \frac{ 1 +\bar{\alpha}^{(2)}}{2}  \}$ (cf. \eqref{eq:bardelta}), the rate and power allocation in~\eqref{eq:powerde},\eqref{eq:powerde1},\eqref{eq:powerde2} (as this policy is explicitly described in the appendix of Section~\ref{sec:solutionpower}) tells us that, the total amount of information, for user 1, in the \emph{private symbols} of a certain phase $s<S$, is equal to
\beq \label{eq:PrivateAmountUser1}
\bigl(\bar{\delta} + ( \bar{\delta} -  \bar{\alpha}^{(2)})^{+}\bigr) T \log P\eeq
bits, while for user 2 this is
\beq \label{eq:PrivateAmountUser2}
\bigl(\bar{\delta} + ( \bar{\delta} -  \bar{\alpha}^{(1)})^{+}\bigr) T \log P\eeq
bits.

The next step is to see how much interference there is to load onto common symbols. Given the power and rate allocation in \eqref{eq:powerde},\eqref{eq:powerde1},\eqref{eq:powerde2},\eqref{eq:bardelta}, it is guaranteed that the accumulated quantized interference in a phase $s<S$ (cf.~\eqref{eq:bitquan}) has $\bigl( (\bar{\delta} - \bar{\alpha}^{(1)})^{+} + ( \bar{\delta} - \bar{\alpha}^{(2)})^{+} \bigr) T \log P$ bits, which can be carried by the common symbols of the next phase $(s+1)$ since they can carry a total of $\bigl(1- \bar{\delta}\bigr)T \log P$ bits (cf.~\eqref{eq:ratejointc}).  This leaves an extra space of $\Delta_{\text{com}} T \log P$ bits in the common symbols, where
\begin{align}
 \Delta_{\text{com}} &\defeq  1- \bar{\delta} - ( \bar{\delta} - \bar{\alpha}^{(1)})^{+} - ( \bar{\delta} - \bar{\alpha}^{(2)})^{+}   \label{eq:com_inter}
\end{align}
is guaranteed to be non-negative for any given $ \bar{\delta}  \leq	\min \{ \bar{\beta}^{(1)},  \bar{\beta}^{(2)}, \frac{ 1 + \bar{\alpha}^{(1)} +  \bar{\alpha}^{(2)}}{3}, \frac{ 1 +\bar{\alpha}^{(2)}}{2}  \}$. This extra space can be split between the two users, by allocating $\omega \Delta_{\text{com}} T \log P$ bits for the message of user 1, and the remaining $(1-\omega) \Delta_{\text{com}} T \log P$ bits for the message of user~2, for some $\omega\in[0,1]$.

Consequently the above, combined with the information stored in private symbols (cf.~\eqref{eq:PrivateAmountUser1},\eqref{eq:PrivateAmountUser2}), allows for
\begin{align}
d_1 &= \bar{\delta} + ( \bar{\delta} -  \bar{\alpha}^{(2)})^{+} +\omega\Delta_{\text{com}}  \label{eq:DoFd1}\\
d_2 &= \bar{\delta} + ( \bar{\delta} -  \bar{\alpha}^{(1)})^{+}  +(1-\omega)\Delta_{\text{com}}.  \label{eq:DoFd2}
\end{align}
The above considers that $S$ is large, and thus removes the effect of having a last phase that carries less new message information.
In the following, we will achieve different corner points by accordingly setting the value of $\omega\in[0,1]$ and of $\bar{\delta}  \leq	\min \{ \bar{\beta}^{(1)},  \bar{\beta}^{(2)}, \frac{ 1 + \bar{\alpha}^{(1)} +  \bar{\alpha}^{(2)}}{3}, \frac{ 1 +\bar{\alpha}^{(2)}}{2}  \}$.

\subsubsection{DoF corner points in Theorem~\ref{thm:genCSIT}}

To achieved the DoF region in Theorem~\ref{thm:genCSIT}, we will show how to achieve the following DoF corner points (see also Table~\ref{tab:outerBoundCornerPoints})
\begin{align}
A &= \bigl(1, \frac{ 1 + \bar{\alpha}^{(2)}}{2} \bigr)  \label{eq:DoFCornerA}\\
B &= \bigl(\bar{\alpha}^{(2)}, 1 \bigr)                 \label{eq:DoFCornerB}\\
C &= \bigl(\frac{2+2\bar{\alpha}^{(1)} - \bar{\alpha}^{(2)} }{3},\frac{2+2\bar{\alpha}^{(2)} - \bar{\alpha}^{(1)}}{3} \bigr) \label{eq:DoFCornerC}\\
D &= \bigl(1, \bar{\alpha}^{(1)} \bigr).        \label{eq:DoFCornerD}
\end{align}
To achieve the DoF region of Theorem~\ref{thm:genCSIT} we need sufficiently good (but certainly not perfect) delayed CSIT such that
\begin{align} \label{eq:betaSuf}
\min \{ \bar{\beta}^{(1)}, \bar{\beta}^{(2)}\} \geq \min \{\frac{1+\bar{\alpha}^{(1)}+\bar{\alpha}^{(2)}}{3}, \frac{1+\bar{\alpha}^{(2)}}{2}\}
\end{align}
(cf. Theorem~\ref{thm:genCSIT}), which in turn implies that  (cf.~\eqref{eq:bardelta})
\[\bar{\delta} \leq \min \{ \frac{ 1 + \bar{\alpha}^{(1)} +  \bar{\alpha}^{(2)}}{3},  \frac{1+  \bar{\alpha}^{(2)}}{2} \} .\]

Under the condition of \eqref{eq:betaSuf}, the DoF corner points are achievable by setting the value of $\omega\in[0,1]$ and of $\bar{\delta}  \leq	\min \{ \frac{ 1 + \bar{\alpha}^{(1)} +  \bar{\alpha}^{(2)}}{3}, \frac{ 1 +\bar{\alpha}^{(2)}}{2}  \}$  as in Table~\ref{tab:outerBoundCornerPoints}.

Specifically when \eqref{eq:betaSuf} and \eqref{eq:case1} hold, we achieve DoF point $B$ by setting $\omega= 0,\bar{\delta} = \bar{\alpha}^{(2)}$ which indeed gives (cf.~\eqref{eq:com_inter},\eqref{eq:DoFd1},\eqref{eq:DoFd2})
\begin{align*}
d_1 &= \bar{\delta} + ( \bar{\delta} -  \bar{\alpha}^{(2)})^{+}  = \bar{\alpha}^{(2)}   \\
d_2 &= \bar{\delta} + ( \bar{\delta} -  \bar{\alpha}^{(1)})^{+}  +  \Delta_{\text{com}} = \bar{\alpha}^{(2)}  +   1 - \bar{\alpha}^{(2)}  =  1.
\end{align*}
To achieve DoF point $D$ we set $\omega= 1$ and $\bar{\delta} = \bar{\alpha}^{(1)}$ and get
\begin{align*}
d_1 &= \bar{\delta} + ( \bar{\delta} -  \bar{\alpha}^{(2)})^{+} + \Delta_{\text{com}} = \bar{\alpha}^{(1)}   +  1- \bar{\alpha}^{(1)}=1\\
d_2 &= \bar{\delta} + ( \bar{\delta} -  \bar{\alpha}^{(1)})^{+}     = \bar{\alpha}^{(1)}
\end{align*}
while to achieve DoF point $C$ we set $\omega= 0$ and $\bar{\delta} = \frac{ 1 + \bar{\alpha}^{(1)} +  \bar{\alpha}^{(2)}}{3}$ and get
\begin{align*}
d_1 &= \bar{\delta} + ( \bar{\delta} -  \bar{\alpha}^{(2)})^{+}  = \frac{2+2\bar{\alpha}^{(1)} - \bar{\alpha}^{(2)} }{3} \\
d_2 &= \bar{\delta} + ( \bar{\delta} -  \bar{\alpha}^{(1)})^{+}  + \Delta_{\text{com}}   = \frac{2+2\bar{\alpha}^{(2)} - \bar{\alpha}^{(1)}}{3}.
\end{align*}

On the other hand, when \eqref{eq:case2} (case 2) and \eqref{eq:betaSuf} hold, to achieve DoF point $B$ we set $\omega= 0$ and $\bar{\delta} = \bar{\alpha}^{(2)}$ as before, while to achieve DoF point $A$, we set $\omega= 0$ and $\bar{\delta} =  \frac{ 1  +  \bar{\alpha}^{(2)}}{2}$.

Finally the entire DoF region of Theorem~\ref{thm:genCSIT} is achieved using time sharing between these corner points.

\begin{table}
\caption{Optimal corner points summary, for sufficiently good delayed CSIT such that $\min \{ \bar{\beta}^{(1)}, \bar{\beta}^{(2)}\} \geq \min \{\frac{1+\bar{\alpha}^{(1)}+\bar{\alpha}^{(2)}}{3}, \frac{1 +   \bar{\alpha}^{(2)}}{2}
\} $.}
\begin{center}
{\renewcommand{\arraystretch}{1.7}
\begin{tabular}{|c|c|c|c|}
  \hline
  Cases            & Corner points  & $\bar{\delta}$          &   $\omega$\\
   \hline
   Case~1          & $C$            &  $ \frac{ 1 + \bar{\alpha}^{(1)} +  \bar{\alpha}^{(2)}}{3}$ &    0  \\
	\cline{2-4}
         & $D$            &  $\bar{\alpha}^{(1)}$   &    1 \\
  \cline{2-4}
           & $B$            &  $\bar{\alpha}^{(2)}$   &    0  \\
				  \hline
   Case~2          & $B$  &  $\bar{\alpha}^{(2)}$    &   0   \\
    \cline{2-4}
									& $A $  &  $ \frac{ 1  +  \bar{\alpha}^{(2)}}{2}$     &  0 \\
			\hline
\end{tabular}
}
\end{center}
\label{tab:outerBoundCornerPoints}
\end{table}

\subsubsection{DoF corner points of Proposition~\ref{prop:genCSITInner}}

\begin{table}
\caption{DoF inner bound corner points, for  delayed CSIT such that $\min \{ \bar{\beta}^{(1)}, \bar{\beta}^{(2)}\} < \min \{\frac{1+\bar{\alpha}^{(1)}+\bar{\alpha}^{(2)}}{3}, \frac{1 +   \bar{\alpha}^{(2)}}{2}
\} $.}
\begin{center}
{\renewcommand{\arraystretch}{1.7}
\begin{tabular}{|c|c|}
  \hline
  Cases            & Corner points  \\
   \hline
     Case~1 and   case of $\min \{ \bar{\beta}^{(1)}, \bar{\beta}^{(2)}\}  \geq  \bar{\alpha}^{(1)} $       & $E, F, B, D$            \\
				  \hline
     Case~1 and   case of $\min \{ \bar{\beta}^{(1)}, \bar{\beta}^{(2)}\}  <  \bar{\alpha}^{(1)} $       & $ B, E, G$              \\
				  \hline
   Case~2          & $B, E, G$     \\
			\hline
\end{tabular}
}
\end{center}
\label{tab:InnerBoundCornerPoints}
\end{table}

Now we focus on the DoF points of Proposition~\ref{prop:genCSITInner}  (see Table~\ref{tab:InnerBoundCornerPoints}). These are the points we label as DoF points $B$ and $D$, as these were defined in \eqref{eq:DoFCornerB} and \eqref{eq:DoFCornerD}, as well as three new DoF points
\begin{align}
E &= \bigl(2\min \{ \bar{\beta}^{(1)}, \bar{\beta}^{(2)}\} - \bar{\alpha}^{(2)}, 1+\bar{\alpha}^{(2)}-\min \{ \bar{\beta}^{(1)}, \bar{\beta}^{(2)}\} \bigr)  \label{eq:DoFCornerE}\\
F  &= \bigl(1+\bar{\alpha}^{(1)}-\min \{ \bar{\beta}^{(1)}, \bar{\beta}^{(2)}\}, 2\min \{ \bar{\beta}^{(1)}, \bar{\beta}^{(2)}\}-\bar{\alpha}^{(1)} \bigr)              \label{eq:DoFCornerF}\\
G &=\bigl(1, \min \{ \bar{\beta}^{(1)}, \bar{\beta}^{(2)}\} \bigr) \label{eq:DoFCornerG}.
\end{align}
As stated in the proposition, we are interested in the range of reduced-quality delayed CSIT, as this is defined by
\begin{align} \label{eq:betaSufNot}
\min \{ \bar{\beta}^{(1)}, \bar{\beta}^{(2)}\} < \min \{\frac{1+\bar{\alpha}^{(1)}+\bar{\alpha}^{(2)}}{3}, \frac{1+\bar{\alpha}^{(2)}}{2}\}
\end{align}
and which implies that $\bar{\delta} \leq \min \{ \bar{\beta}^{(1)}, \bar{\beta}^{(2)}\}$ (cf.~\eqref{eq:bardelta}).
In addition to the two cases in \eqref{eq:case1},\eqref{eq:case2}, we now additionally consider the cases where
\begin{align}
\min \{ \bar{\beta}^{(1)}, \bar{\beta}^{(2)}\} & \geq  \bar{\alpha}^{(1)} \label{eq:betaAlpha1Suf}\\
\min \{ \bar{\beta}^{(1)}, \bar{\beta}^{(2)}\} & <   \bar{\alpha}^{(1)}.\label{eq:betaAlpha1SufNot}
\end{align}

When \eqref{eq:case1},\eqref{eq:betaSufNot} and \eqref{eq:betaAlpha1Suf} hold, we set $\omega= 0,\bar{\delta} = \bar{\alpha}^{(2)}$ as before to achieve DoF point $B$.  To achieve point $D$, we set $\omega= 1$ and $\bar{\delta} = \bar{\alpha}^{(1)}$ as before, whereas to achieve point $E$, we set $\omega= 0,\bar{\delta} =  \min \{ \bar{\beta}^{(1)}, \bar{\beta}^{(2)}\}$ to get (cf.~\eqref{eq:com_inter},~\eqref{eq:DoFd1},~\eqref{eq:DoFd2})
\begin{align*}
d_1 &= \bar{\delta} + ( \bar{\delta} -  \bar{\alpha}^{(2)})^{+}    =   2\min \{ \bar{\beta}^{(1)}, \bar{\beta}^{(2)}\} - \bar{\alpha}^{(2)}\\
d_2 &= \bar{\delta} + ( \bar{\delta} -  \bar{\alpha}^{(1)})^{+} + \Delta_{\text{com}}  = 1+\bar{\alpha}^{(2)}-\min \{ \bar{\beta}^{(1)}, \bar{\beta}^{(2)}\}.
\end{align*}
Finally to achieve DoF point $F$, we set $\omega =1$ and $\bar{\delta} =  \min \{ \bar{\beta}^{(1)}, \bar{\beta}^{(2)}\}$.

When \eqref{eq:case1},\eqref{eq:betaSufNot} and \eqref{eq:betaAlpha1SufNot} hold, we achieve points $B$ and $E$ with the same parameters as before, while to achieve point $G$, we set $\omega =1,\bar{\delta} =  \min \{ \bar{\beta}^{(1)}, \bar{\beta}^{(2)}\}$.

Similarly when \eqref{eq:case2} and \eqref{eq:betaSufNot} hold, we achieve points $B, E, G$ by setting $\omega$ and $\bar{\delta}$ as above.

Finally the entire DoF region of Proposition~\ref{prop:genCSITInner} is achieved with time sharing between the corner points.

\subsection{Scheme $\Xc$: examples \label{sec:scheme_exam}}
We proceed to provide example instances of our general scheme, for specific cases of particular interest.

\subsubsection{Fixed and imperfect quality delayed CSIT, no current CSIT}
We consider the case of no current CSIT ($\alpha^{(i)}_t=0,\ \forall t, i$) and of imperfect delayed CSIT of an unchanged quality $\beta^{(1)}_t=\beta^{(2)}_t\leq 1,\forall t$.  We focus on the case of $\beta^{(1)}_t=\beta^{(2)}_t = 1/3,\forall t$. The universal scheme - with these parameters - achieves the optimal DoF by achieving the optimal DoF corner point $(d_1=\frac{2}{3}, d_2=\frac{2}{3})$, as in the case of~\cite{MAT:11c} which assumed that the delayed feedback of a channel could be sent with perfect quality.

For this case of $\beta^{(i)}_t=1/3,\alpha^{(i)}_t=0$, we have $\bar{\alpha}^{(1)} =\bar{\alpha}^{(2)} = 0 $, $\bar{\beta}^{(1)} =\bar{\beta}^{(2)} = 1/3$. Toward designing the scheme, we set $\bar{\delta} = 1/3$ (cf.~\eqref{eq:bardelta}). For the case of block fading where we can rewrite the time index to reflect a unit coherence period, delayed CSIT is simply the CSIT that comes during the next coherence period, i.e., during the next time slot.  Given the i.i.d. fast fading assumption (\cite{MAT:11c}), we can set $\eta=1$ (cf.~\eqref{eq:beta1},\eqref{eq:beta2}), which allows for a simpler variant of our scheme where now the phases have duration $T=1$.
In this simplified variant, the transmitted signal (cf.~\eqref{eq:TxX1Ph1}) takes the simple form
\begin{align*}
\xv_{t} = \wv_t c_t +  \Bmatrix{ a_t \\ a^{'}_t} + \Bmatrix{ b_t \\ b^{'}_t} \end{align*}
with the power and rates of the symbols (cf.~\eqref{eq:RPowerX1Ph1}) set as
\begin{equation}
\begin{array}{c}
P^{(c)}_t \doteq  P ,  \     r^{(c)}_t  = 1-1/3 \\
P^{(a)}_t \doteq  P^{(a')}_t \doteq P^{(b)}_t \doteq  P^{(b')}_t \doteq  P^{1/3}\\
r^{(a)}_t   = r^{(a')}_t =r^{(b)}_t   = r^{(b')}_t = 1/3 . \end{array}
\end{equation}
During each phase, the transmitter quantizes - as instructed in \eqref{eq:bitquan} - the interference accumulated in that phase, with a quantization rate of $2/3 \log P$, which is mapped into the common symbol $c_{t+1}$ that will be transmitted in the next phase (at time-slot $t+1$). For large enough communication length, simple calculations can show that this can achieve the optimal DoF  $(d_1=\frac{2}{3}, d_2=\frac{2}{3})$, and can do so with imperfect quality CSIT.
Table~\ref{tab:bitsummary-1} summarizes the rates associated to the symbols in this scheme.

\begin{table}
\caption{Bits carried by private symbols, common symbols, and by the quantized interference, for phase~$s=1,2,\cdots,S-1$.}
\begin{center}
\begin{tabular}{|c|c|}
  \hline
                & Total bits ($\log P$)\\
   \hline
   Private symbols for user~1    &  $2/3$ \\
   \hline
   Private symbols for user~2    &  $2/3$ \\
   \hline
   Common symbols                 &  $2/3$  \\
    \hline
  Quantized interference          &  $2/3$  \\
    \hline
\end{tabular}
\end{center}
\label{tab:bitsummary-1}
\end{table}

\subsubsection{Alternating between two current-CSIT states\label{sec:schemeSpecificToAlternating}}

In the context of the two-user MISO BC with spatially and temporally i.i.d. fading and $M=2$, the work in~\cite{TJSP:12} considered the \emph{alternating CSIT} setting where CSIT for the two users, alternates between perfect current CSIT (labeled here as state $P$), perfect delayed CSIT ($D$), or no CSIT ($N$).
In this setting where $I_i$ denoted the CSIT state for the channel of user~$i$ at any given time ($I_1, I_2 \in \{P,D,N\}$), the work in~\cite{TJSP:12} considered communication where, for a fraction $\lambda_{I_1I_2}$ of the time, the CSIT states are equal to $I_1,I_2$ (state $I_1$ for the first user, state $I_2$ for the second user).  The same work focused on the symmetric case where $\lambda_{I_1I_2}=\lambda_{I_2I_1}$.  For $\lambda_P \defeq \sum_{I_2 \in \{P,D,N\}}\lambda_{PI_2}$ being the fraction of the time where one user has perfect CSIT, and $\lambda_D\defeq \sum_{I_2 \in \{P,D,N\}}\lambda_{DI_2}$ being the fraction of the time where one user had delayed CSIT, the work in \cite{TJSP:12} characterized the optimal DoF region to take the form
\begin{align*}
d_1\leq 1&,\ \ d_1\leq 1, \\
d_1+2d_2&\leq 2+\lambda_P\\
d_2+2d_1&\leq 2+\lambda_P\\
d_1+d_2&\leq 1+\lambda_P+\lambda_D.
\end{align*}
The above setting corresponds to our symmetric setting where $\alpha_t^{(1)}, \beta_t^{(1)}, \alpha_t^{(2)}, \beta_t^{(2)} \in \{0,1\}$, $\forall t$, and where \bea
\lambda_P &=& \bar{\alpha}^{(1)}=\bar{\alpha}^{(2)}\\
\lambda_D &=& \bar{\beta}^{(1)}-\bar{\alpha}^{(1)}=\bar{\beta}^{(2)}-\bar{\alpha}^{(2)}
\eea
in which case our DoF inner bound matches the above, and as a result, for any $\bar{\beta}\geq \frac{1+2\bar{\alpha}}{3}$, Theorem~1 generalizes~\cite{TJSP:12} to any set of quality exponents, avoiding the symmetry assumption, as well as easing on the i.i.d. block-fading assumption.

The universal scheme described in this section, can be directly applied to optimally implement more general alternating CSIT settings.
We here offer an example where, in the presence of sufficiently good delayed CSIT, the current CSIT of the two users alternates between two quality exponents equal to $\frac{1}{2}$ and $\frac{3}{4}$, i.e.,
\beqn
\begin{array}{cccccc}
                 &  t=1          &     t=2     &      t=3    & t=4          & \cdots\\
\alpha^{(1)}_t = &  \frac{1}{2}  & \frac{3}{4} &\frac{1}{2}  & \frac{3}{4}  & \cdots\\
\alpha^{(2)}_t = &  \frac{3}{4}  & \frac{1}{2} & \frac{3}{4} & \frac{1}{2}  & \cdots
\end{array} \eeqn
In this case, which corresponds to having $\bar{\alpha}^{(1)} =\bar{\alpha}^{(2)} = 5/8 $, we can choose any delayed CSIT process that gives $\bar{\beta}^{(1)} =\bar{\beta}^{(2)} = 3/4$ which suffices (see Corollary~\ref{cor:ImperfectVsPerfectDelayed}) to achieve the optimal DoF region by achieving the optimal DoF point $(d_1=\frac{7}{8}, d_2=\frac{7}{8})$.

Toward designing the scheme, we set $\bar{\delta} = 3/4$.  For this example, and again considering a block-fading fast-fading setting (unit-length coherence period), the scheme can have phases with duration $T=2$. The transmitted signal (cf.~\eqref{eq:TxX1Ph1}) now takes the form
\begin{align*}
\xv_{t} = \wv_t c_t+\hat{\gv}^{\bot}_t a_t + \hat{\hv}_t a^{'}_t + \hat{\hv}^{\bot}_t b_t + \hat{\gv}_t b^{'}_t \end{align*}
with power and rates of the symbols being set as instructed in~\eqref{eq:RPowerX1Ph1}.
Again as instructed by the general description of the scheme, at the end of  phase~$s=1,2,\cdots,S-1$, the transmitter quantizes the interference accumulated during that phase, and does so using a total of $2(1/8 +1/8) \log P$ quantization bits (cf.~\eqref{eq:bitquan}).  These bits are then mapped into the common symbols that will be transmitted in the next phase.
For a large number of phases, the proposed scheme achieves the optimal DoF point $(d_1=\frac{7}{8}, d_2=\frac{7}{8})$.
Table~\ref{tab:bitsummary-2} summarizes the rates associated to the symbols in this scheme.

\begin{table}
\caption{Bits carried by private symbols, common symbols, and by the quantized interference, for phase~$s$, $s=1,2,\cdots,S-1$, of the alternating CSIT scheme.}
\begin{center}
\begin{tabular}{|c|c|}
  \hline
                & Total bits ($\times \log P$)\\
   \hline
   Private symbols for user~1    &  $(7\times 2)/8$ \\
   \hline
   Private symbols for user~2    &  $(7\times 2)/8$ \\
   \hline
   Common symbols                 &  $(1\times 2)/4$  \\
    \hline
  Quantized interference          &  $(1\times 2)/4$  \\
    \hline
\end{tabular}
\end{center}
\label{tab:bitsummary-2}
\end{table}

\subsubsection{Schemes with short duration\label{sec:schemeSpecificFiniteBlock}}
We recall that the Maddah-Ali and Tse scheme~\cite{MAT:11c} uses (under the employed assumption in~\cite{MAT:11c} of a unit coherence period) $T=3$ channel uses, during which it employs\footnote{We here refer to an equivalent MAT scheme that can be seen as a special case of the scheme in \cite{YKGY:12d} for $\alpha = 0$.} $\beta^{(1)}_1=1,\beta^{(2)}_1=1$ (the rest of the exponents are zero). The scheme manages to have the information bits of the quantized interference, `fit' inside the common symbols in the above three time slots.

A similar setting where again the information bits of the quantized interference, can fit in the common symbols of a single, short phase, would be if
\beqn
\begin{array}{cccccc}
                 &  t=1          &     t=2     &      t=3    & t=4          & \cdots\\
\alpha^{(1)}_t = &  0           & 0            &\frac{1}{4}  & 0            & \cdots\\
\beta^{(1)}_t = &  1            & \frac{1}{4} & \frac{1}{4} &  0            & \cdots\\
\alpha^{(2)}_t = &  0           & \frac{1}{4} &    0        &    0          & \cdots\\
\beta^{(2)}_t = &      1        & \frac{1}{4} & \frac{1}{4} &     0         & \cdots\\
\end{array} \eeqn
where the corresponding single-phase ($T=4$ time-slots) scheme, can achieve the optimal DoF corner point $(d_1=\frac{11}{16}, d_2=\frac{11}{16})$.

\section{Conclusions} \label{sec:conclu}
The work made progress toward establishing and meeting the limits of using imperfect and delayed feedback. Considering a general CSIT process and a primitive measure of CSIT quality, the work provided DoF expressions that are simple and insightful functions of easy to calculate parameters which concisely capture the problem complexity. The derived insight addresses practical questions on topics relating to the usefulness of predicted, current and delayed CSIT, the impact of estimate precision, the effect of feedback delays, and the benefit of having feedback symmetry by employing comparable feedback links across users.
Further insight was derived from the introduced \emph{periodically evolving feedback} setting, which captures many of the engineering options in practical feedback settings.

In terms of the applicability of the DoF high-SNR asymptotic approach, for our chosen setting of a small number of users (two in this case), we expect the high-SNR insights to hold for SNR values of operational interest. The nature of the improved bounds and novel constructions, allows for this same insight to hold for a broad family of block fading and non-block fading channel models.


We believe that the adopted approach is fundamental, in the sense that it considers a general fading process, a general CSIT process, and a primitive measure of feedback quality in the form of the precision of estimates at any time about any channel, i.e., in the form of the entire set of estimation errors $\{(\bfh_t-\hat{\bfh}_{t,t'}),(\bfg_t-\hat{\bfg}_{t,t'})\}_{t,t' = 1}^n$ at any time about any channel.
As we have seen, this set of errors naturally fluctuates depending on the instance of the problem, and as expected, the overall optimal performance is defined by the statistics of this error set. These statistics are mildly constrained to the case of having Gaussian estimation errors which are independent of the prior and current channel estimates\footnote{Again we caution the reader that this is not an assumption about independence between errors, but rather between errors and estimates.}. Under these assumptions, the results capture the performance effect of the statistics of feedback.  Interestingly this effect - at least for sufficiently good delayed CSIT, and for high SNR - is captured by the averages of the quality exponents. As noted, this can be traced back to the assumption that the estimation errors are Gaussian, which means that the statistics of $\{(\bfh_t-\hat{\bfh}_{t,t'}),(\bfg_t-\hat{\bfg}_{t,t'})\}_{t,t' = 1}^n$ are captured by a covariance matrix that has diagonal (block) entries of the form
$\{ \frac{1}{M}\E[||\bfh_t-\hat{\bfh}_{t,t'}||_{F}^2], \frac{1}{M}\E[||\bfg_t-\hat{\bfg}_{t,t'}||_{F}^2]\}_{t,t' = 1}^n$, and whose off-diagonal entries are not used by the scheme, but where this scheme though meets an outer bound that has kept open the possibility of any off-diagonal elements. Hence, as stated, under our assumptions, the essence of the CSIT error statistics is captured by the diagonal block elements (of the aforementioned covariance matrix) whose effects are in turn captured -  in the high-SNR regime - by the quality exponents.

This general approach allows for consideration of many facets of the performance-vs-feedback question in the two-user MISO BC setting, accentuating some important facets while revealing the reduced role of other facets.  For example, while the approach allows for consideration of predicted CSIT - i.e., of estimates for future channels - the result at the end reveals that such estimates do not provide DoF gains, again under our assumptions. In a similar manner, the result leaves open the possibility of a role in the off-diagonal elements of the aforementioned covariance matrix of estimation errors, but in the end again reveals that these can be neglected without a DoF effect. Similarly, the approach allows for any `typical' sequence of quality exponents - thus avoiding the need to assume periodic or static feedback processes or a block-fading structure - but despite this generality in the range of the considered exponents, in the end the result reveals that what really matters is the long-term average of each of these sequences of current and delayed CSIT exponents.

Finally we believe the main assumptions here to be mild.
Regarding the high SNR assumption, there is substantial evidence that for primitive networks (such as the BC and the IC) with a reasonably small number of users, DoF analysis offers good insight on the performance at moderate SNR.
Any possible extensions though to the setting of larger cellular networks, may need to consider saturation effects on the high-SNR spectral efficiency, as these were recently revealed in~\cite{LHA:13} to hold for settings where communication involves clusters of large size.
Furthermore the assumption of having global CSIR, allowed us to focus on the question of feedback to the transmitters, which is a fundamental question on its own. While the overhead of gathering global CSIR must not be neglected, it has been repeatedly shown (cf.~\cite{APRC:11,KC:12}) that this overhead is manageable in the presence of a reduced number of users.  When considering extensions to other multiuser networks with potentially more users, such analysis may have to be combined with finding ways to disseminate imperfect global CSIR (cf.~\cite{APRC:11,KC:12,CE:13spawc}, see also~\cite{LSY:13,ALH:12}) whose effect increases as the number of users increases. Additionally asking that current estimation errors are independent of current estimates, is a widely accepted assumption.  Similarly accepted is the assumption that the estimation error is independent of the past estimates, as this assumption suggests good feedback processes that utilize possible correlations to improve current channel estimates. Finally the requirement that the running average of the quality exponents of a single user, converges to a fixed value after a sufficiently long time, is also believed to be reasonable, as it would hold even if these exponents were themselves treated as random variables from an ergodic process.

\section{Appendix - Proof of outer bound Lemma \label{sec:outerb}}
\begin{proof}
Let $W_1,W_2$ respectively denote the messages for the first and second user, and let $R_1,R_2$ denote the two users' rates.  Each user sends their message over $n$ channel uses, where $n$ is large.  For ease of exposition we introduce the following notation.
\begin{align*}
\Sm_t \defeq & \ \Bmatrix{\hv^\T_t \\ \gv^\T_t}, \quad \check{\Sm}_t \defeq \Bmatrix{\check{\hv}^\T_t \\ \check{\gv}^\T_t}, \quad  \hat{\Sm}_t \defeq \Bmatrix{\hat{\hv}^\T_t \\ \hat{\gv}^\T_t}, \quad \zv_t \defeq  \Bmatrix{z_t^{(1)} \\ z_t^{(2)}}  \\
y^{(i)}_{[n]} \defeq &  \  \{y^{(i)}_{t}\}_{t=1}^{n},\quad i=1,2 \\
\Omega_{[n]} \defeq &  \ \{ \Sm_t,\check{\Sm}_t, \hat{\Sm}_t \}_{t=1}^{n}.
\end{align*}

The first step is to construct a degraded BC by providing the first user with complete and immediately available information on the second user's received signal.  In this improved scenario, the following bounds hold.
\begin{align}
& nR_1 \nonumber\\
& = H(W_1) \nonumber\\
&= H(W_1|\Omega_{[n]}) \nonumber\\
&\leq I(W_1;y^{(1)}_{[n]},y^{(2)}_{[n]}|\Omega_{[n]}) + n \epsilon_n \label{eq:R1b0}\\
&\leq I(W_1;W_2,y^{(1)}_{[n]},y^{(2)}_{[n]}|\Omega_{[n]}) + n \epsilon_n \nonumber\\
&= I(W_1;y^{(1)}_{[n]},y^{(2)}_{[n]}|W_2,\Omega_{[n]}) + n \epsilon_n \nonumber\\
&= h(y^{(1)}_{[n]},y^{(2)}_{[n]}|W_2,\Omega_{[n]}) - \underbrace{h(y^{(1)}_{[n]},y^{(2)}_{[n]}|W_1, W_2,\Omega_{[n]})}_{no(\log P)} + n \epsilon_n     \nonumber\\
&= \sum^{n}_{t=1}h(y^{(1)}_t,y^{(2)}_t|y^{(1)}_{[t-1]},y^{(2)}_{[t-1]}, W_2,\Omega_{[n]})  + no(\log P) +n \epsilon_n   \label{eq:R1b3}
\end{align}
where \eqref{eq:R1b0} results from Fano's inequality, where $y^{(i)}_{0}$ was set to zero by convention, and where the last equality follows from the entropy chain rule and the fact that the knowledge of $\{W_1, W_2,\Omega_{[n]}\}$ implies knowledge of $\{y^{(1)}_{[n]},y^{(2)}_{[n]}\}$ up to noise level.

Similarly
\begin{align}
&nR_2 \nonumber\\
&= H(W_2) \nonumber\\
&\leq  I(W_2;y^{(2)}_{[n]}| \Omega_{[n]}) + n \epsilon_n  \label{eq:R2b0}\\
& =  \underbrace{h(y^{(2)}_{[n]}| \Omega_{[n]})}_{ \leq n\log P + n o(\log P) } - h(y^{(2)}_{[n]}| W_2 ,\Omega_{[n]}) + n \epsilon_n  \label{eq:R2b-power}\\
& \leq - \!\sum^{n}_{t=1} h(y^{(2)}_t|y^{(2)}_{[t-1]},W_2,\Omega_{[n]})  \!+ \!n\log P \!+\! n o(\log P)   \!+ n \epsilon_n  \label{eq:R2b-chain}\\
& \leq - \sum^{n}_{t=1} h(y^{(2)}_t|y^{(1)}_{[t-1]}, y^{(2)}_{[t-1]},W_2,\Omega_{[n]})  \nonumber\\ & 	 \quad\quad\quad\quad\quad  + n\log P  + n o(\log P)   + n \epsilon_n  \label{eq:R2b3}
\end{align}
where \eqref{eq:R2b-chain} follows from the entropy chain rule and from the fact that received signals are scalars, while the last step is due to the fact that conditioning reduces entropy.

Now given \eqref{eq:R1b3} and \eqref{eq:R2b3}, we upper bound $R_1+2R_2$ as
\begin{align}
&n(R_1+2R_2) \leq  2n\log P  + n o(\log P) + 3n\epsilon_n   \nonumber\\
&  + \sum^{n}_{t=1} \left(h(y^{(1)}_t,y^{(2)}_t|  U, S_t,\hat{S}_t)  -2h(y^{(2)}_t|U, S_t,\hat{S}_t)\right)     \label{eq:R12b}
\end{align}
where \[U \defeq \{y^{(1)}_{[t-1]},y^{(2)}_{[t-1]}, W_2,\Omega_{[n]}\} \setminus  S_t,\hat{S}_t \]
and where each term $h(y^{(1)}_t,y^{(2)}_t|  U, S_t,\hat{S}_t)  -2h(y^{(2)}_t|U, S_t,\hat{S}_t)$ in the summation, can be upper bounded as
\begin{align}
&h(y^{(1)}_t,y^{(2)}_t|U, S_t,\hat{S}_t)  -2h(y^{(2)}_t|U, S_t,\hat{S}_t) \nonumber\\
& \leq  \max_{\stackrel{P_{X_t}}{\E[\text{tr}(X_t X_t^\H)]\le P}}  \Bigl [ h(y^{(1)}_t,y^{(2)}_t|U, S_t,\hat{S}_t)-2h(y^{(2)}_t|U, S_t,\hat{S}_t) \Bigr] \nonumber\\
& \leq  \E_{\hat{S}_t}\!\! \max_{ \stackrel{P_{X_t}}{\E[\text{tr}(X_t X_t^\H)]\le  P } } \!\!\! \E_{S_t|\hat{S}_t}   \Bigl [ h(y^{(1)}_t,y^{(2)}_t|U, S_t = \Sm_t , \hat{S}_t=\hat{\Sm}_t )  \nonumber\\
& \quad\quad\quad\quad\quad\quad\quad\quad\quad\quad\quad \ - 2h(y^{(2)}_t|U, S_t = \Sm_t , \hat{S}_t=\hat{\Sm}_t  ) \Bigr]\nonumber\\
& =  \E_{\hat{S}_t}\!\!\!\!  \max_{ \stackrel{P_{X_t}}{\E[\text{tr}(X_t X_t^\H)]\le  P } }  \!\!\!\!\! \E_{\tilde{S}_t}  \Bigl [ h( \Sm_t \xv_t + \zv_t|U) - 2h( \gv_t^\T \xv_t + z_t^{(2)}|U) \Bigr]  \nonumber\\
& = \! \E_{\hat{S}_t}\! \max_{\Psim \succeq 0: \text{tr}(\Psim)\le  P}  \!\!\!\E_{\tilde{S}_t}  \! \bigl[ \log \det\!  \left( I \!\!+ \! \Sm_t\Psim  \Sm_t^\H \right) \!-\! 2  \log  \left( 1 \!\!+\!  \gv_t^\H \Psim  \gv_t \right) \Bigr] \label{eq:R12b-opt} \\
&\leq \E_{\hat{S}_t}\max_{\Psim \succeq 0: \text{tr}(\Psim)\le  P} \E_{\tilde{S}_t}   \bigl[ \log \left( 1 +  \hv_t^\H \Psim  \hv_t \right) -  \log  \left( 1 +  \gv_t^\H \Psim  \gv_t \right) \Bigr].  \label{eq:R12b1}
\end{align}
In the above, \eqref{eq:R12b-opt} uses the results in \cite[Corollary~4]{WLS+:09} that tell us that Gaussian input maximizes the weighted difference of two differential entropies\footnote{We note that the results in \cite[Corollary~4]{WLS+:09} are described for the non-fading channel model, however, as argued in the same work in \cite[Section~V]{WLS+:09}, the results can be readily extended to the fading channel model by linearly transforming the fading channel into an equivalent non-fading channel, with the new channel actually maintaining the same capacity and the same degradedness order.}, as long as: 1) $y^{(2)}_t$ is a degraded version of $\{y^{(1)}_t,y^{(2)}_t\}$;  2) $U$ is independent of $z_t^{(1)},z_t^{(2)}$; 3) the input maximization is done given a fixed fading realization $\hat{\Sm}_t$, and is independent of $\tilde{\Sm}_t$ \footnote{We recall that $\xv_t$ is only a function of the messages and of the CSIT (current and delayed) estimates up to time $t$, and that these CSIT estimates are assumed to be independent of the current estimate errors at time $t$.}. Furthermore, in the above, \eqref{eq:R12b1} comes from Fischer's inequality which gives that $\det (I +  \Sm_t\Psim  \Sm_t^\H ) \leq (1 +\hv_t^\H \Psim \hv_t)(1 +\gv_t^\H \Psim  \gv_t)$.

At this point we follow the steps involving equation (25) in \cite{YKGY:12d}, to upper bound the right hand side of~\eqref{eq:R12b1} as
\begin{align}
&\E_{\hat{S}_t}\max_{\Psim \succeq 0: \text{tr}(\Psim)\le  P} \E_{\tilde{S}_t}   \bigl[ \log \left( 1 +  \hv_t^\H \Psim  \hv_t \right) -  \log  \left( 1 +  \gv_t^\H \Psim  \gv_t \right) \Bigr] \nonumber\\
&\leq \alpha^{(2)}_t\log P+o(\log P).   \label{eq:ubsheng}
\end{align}
Combining~\eqref{eq:R12b} and~\eqref{eq:R12b1}, gives that
$n(R_1+2R_2) \leq \sum^{n}_{t=1}  \left((2+\alpha^{(2)}_t)\log P+ o(\log P)+ 3 \epsilon_n \right)$
and consequently that
\beqn
d_1+2d_2 \leq  2 +\bar{\alpha}^{(2)}.     \label{eq:R12bf}
\eeqn
Similarly, interchanging the roles of the two users, allows for
\beqn
d_2+2d_1 \leq 2+\bar{\alpha}^{(1)}.
\eeqn
Finally the fact that each user has a single receive antenna, gives that $d_1\leq 1, d_2\leq 1$.
\end{proof}

\section{Appendix - Further details on the scheme} \label{sec:DetailsX}

\subsection{Explicit power allocation solutions under constraints in equations \eqref{eq:powerde},\eqref{eq:powerde1},\eqref{eq:powerde2}} \label{sec:solutionpower}
We remind the reader that, in designing the power allocation policy of the scheme, we must design the power parameters $\{\delta_t^{(1)}, \delta_t^{(2)}\}_{t\in\Bc_s}$ to satisfy equations \eqref{eq:powerde},\eqref{eq:powerde1},\eqref{eq:powerde2} which asked that
\begin{align*}
\beta_t^{(i)} & \geq \delta_t^{(i)}  \quad  i=1,2, \  t \in \Bc_s  \\
\frac{1}{T}\sum_{t\in \Bc_s}\delta_t^{(1)}  & =\frac{1}{T}\sum_{t\in \Bc_s}\delta_t^{(2)} = \bar{\delta}   \\
\frac{1}{T}\sum_{t\in \Bc_s} (\delta_t^{(i)}- \alpha_t^{(i)})^{+}  & =(  \bar{\delta} - \bar{\alpha}^{(i)})^{+} \quad i=1,2
\end{align*}
for a given $\bar{\delta} \in [0, 1]$.
For each phase $s$, we here explicitly describe such sequence $\{\delta_t^{(1)}, \delta_t^{(2)}\}_{t\in\Bc_s}$, which is constructed using a waterfilling-like approach.

We first consider the case where $\bar{\delta}  \geq \bar{\alpha}^{(i)}.$
At any given time $t = \Bc_{s,1}, \Bc_{s,2}, \cdots, \Bc_{s,T}$, we set
\begin{align*}  
\delta_{t}^{(i)} \!\!=\!\! \left \{ \begin{array}{c}
      \!\!\! T (\bar{\delta} \!-\! \bar{\alpha}^{(i)}) \!-\! \Delta_{\delta,t}  \!+ \!  \alpha_{t}^{(i)}   \ \text{if} \  \beta_{t}^{(i)}  \!\geq\!  T (\bar{\delta} \!-\! \bar{\alpha}^{(i)}) \!-\!\Delta_{\delta,t} \!+\! \alpha_{t}^{(i)}   \\
     \beta_{t}^{(i)} \quad \quad\quad\quad \quad\quad \quad  \text{if} \  \beta_{t}^{(i)}  \!<\!  T (\bar{\delta} \!-\! \bar{\alpha}^{(i)}) \!-\!\Delta_{\delta,t} \!+\! \alpha_{t}^{(i)}
      \end{array}
\right.
\end{align*}
where $\Delta_{\delta,t}$ is initialized to zero ($\Delta_{\delta,\Bc_{s,1}} = 0$), and is updated each time, so that the calculation of $\delta_{t+1}^{(i)}$, uses
\[\Delta_{\delta,t+1} = \Delta_{\delta,t} + \delta_{t}^{(i)} - \alpha_{t}^{(i)}. \]
In the end, the solution takes the form
\begin{align*}  
\delta_{t}^{(i)} \!\!=\!\! \left \{ \begin{array}{c}
      \beta_{t}^{(i)},  \quad \quad \quad\quad \quad \quad\quad \quad \quad     t=\Bc_{s,1}, \cdots, \Bc_{s,\tau'-1} \\
     \!\!\! T (\bar{\delta} \!-\! \bar{\alpha}^{(i)}) \!+\!   \alpha_{t}^{(i)} \!-\! \sum_{\ell=1}^{\tau'-1}  (\beta_{\Bc_{s,\ell} }^{(i)} \!-\! \alpha_{\Bc_{s,\ell}}^{(i)} ),    \quad  t= \Bc_{s,\tau'}  \\
		\alpha_{t}^{(i)},  \quad \quad \quad\quad \quad \quad\quad \quad \quad    t=\Bc_{s,\tau'+1}, \cdots, \Bc_{s,T}
      \end{array}
\right.
\end{align*}
where $\tau'$ is a function\footnote{Note that there is no need to explicitly describe $\tau'$, because the schemes are explicitly described as a function of the above $\delta_{t}^{(i)}$, which - after calculation - also reveal $\tau'$ which - by design - falls within the proper range.} of the quality exponents during phase $s$.  This design of $\{\delta_t^{(i)}\}_{t\in\Bc_s}$ satisfies \eqref{eq:powerde},\eqref{eq:powerde1}, as well as \eqref{eq:powerde2}, since, for the case where $\bar{\delta} - \bar{\alpha}^{(i)} \geq 0$, we deliberately force $\delta_{t}^{(i)} - \alpha_{t}^{(i)} \geq 0 , \ t \in \Bc_s. $

Similarly for $\bar{\delta} \leq \bar{\alpha}^{(i)}$, we set
\begin{align*}
\delta_{t}^{(i)} = \left \{ \begin{array}{c}
        \alpha_{t}^{(i)}       \quad \quad\quad    \text{if} \quad   \alpha_{t}^{(i)}  \leq  T \bar{\delta} - \Delta_{\delta,t}   \\
        T \bar{\delta} - \Delta_{\delta,t}   \quad   \text{if} \quad   \alpha_{t}^{(i)}   >    T \bar{\delta} - \Delta_{\delta,t}
      \end{array}
\right.
\end{align*}
where $\Delta_{\delta,t}$ is initialized to zero, and is updated as
\[\Delta_{\delta,t+1} = \Delta_{\delta,t}  + \delta_{t}^{(i)}. \]
In the end, in this case, the solution takes the form
\begin{align*}
\delta_{t}^{(i)} =\left \{ \begin{array}{c}
      \alpha_{t}^{(i)},   \quad \quad \quad     t=\Bc_{s,1}, \cdots, \Bc_{s,\tau'-1} \\
      T \bar{\delta} - \sum_{\ell=1}^{\tau'-1} \alpha_{\Bc_{s,\ell}}^{(i)},   \quad \quad \quad  t= \Bc_{s,\tau'}  \\
		0, \quad\quad \quad    t=\Bc_{s,\tau'+1}, \cdots, \Bc_{s,T}
      \end{array}
\right.
\end{align*}
where again $\tau'$ is a function of the quality exponents during phase $s$. This satisfies \eqref{eq:powerde},\eqref{eq:powerde1}, as well as \eqref{eq:powerde2} since, for the case where $\bar{\delta} - \bar{\alpha}^{(i)} \leq 0$, we again have $\delta_{t}^{(i)} - \alpha_{t}^{(i)} \leq 0 , \ t\in \Bc_s$.

\subsection{Encoding and decoding details for steps in equations~\eqref{eq:sch1Combits},\eqref{eq:sch1Combitss}} \label{sec:DetailsX2}
We here elaborate on how the users will be able to decode the amount of feedback bits described in equations~\eqref{eq:sch1Combits} and \eqref{eq:sch1Combitss}. We first provide the following lemma, which holds for any $T$.

\vspace{3pt}
\begin{lemma} \label{lem:latticecode}
 Let
  \begin{align}
	    \bar{y}_t^{(1)} &= c_t +  P^{ \frac{\delta^{(2)}_{t}}{2}} \bar{z}_t^{(1)}, \label{eq:latticech}\\
			\bar{y}_t^{(2)} &= c_t +  P^{ \frac{\delta^{(1)}_{t}}{2}} \bar{z}_t^{(2)}, \quad  t = 1,2, \cdots,\! T \label{eq:latticech1}
  \end{align}
	where $\E[ |c_t|^2 ] \le P$,  $Pr (|\bar{z}_t^{(i)}|^2 > P^{\epsilon}) \doteq 0$, and $\frac{1}{T}\sum_{t=1}^{T}\delta^{(i)}_{t}  \leq  \bar{\delta}^{\ast}$ for a given $\bar{\delta}^{\ast}\in [0, 1]$, $i=1,2$.  Also let $r\defeq 1-\bar{\delta}^{\ast}-\epsilon $ for a vanishingly small but positive $\epsilon>0$, and consider communication over $T$ channel uses. Then for any rate up to $R = r\log P -o(\log P)$ (bits/channel use), the probability of error can be made to vanish with asymptotically increasing SNR.
\end{lemma}
\vspace{3pt}

\begin{proof}
We will draw each $T$-length codevector $$\cv \defeq [c_{1}, \cdots, c_{T}]^\T$$ from a lattice code of the form
\begin{align} \label{eq:sch2proofc}
\{\theta \Mm \qv  \ |  \ \qv \in \aleph\}
\end{align}
where $\aleph \subset \C^{T}$ is the $T$-dimensional $2^R$-QAM constellation, where $\Mm \in \C^{T\times T}$ is a specifically constructed unitary matrix of algebraic conjugates that allows for the \emph{non vanishing product distance} property (to be described later on - see for example \cite{BVR+:96}), and where
\begin{align} \label{eq:sch2theta}
\theta = P^{\frac{1-r}{2}} = P^{(\bar{\delta}^{\ast}+\epsilon)/2}
\end{align}
is designed to guarantee that $\E{||\cv||^2} \doteq P$ (to derive this value of $\theta$, just recall the QAM property that $\E{||\qv||^2} \doteq 2^R \doteq P^{r}$).
Specifically for any two codevectors $\cv = [c_{1}, \cdots, c_{T}]^\T, \cv^{\star} = [c^{\star}_{1}, \cdots, c^{\star}_{T}]^\T$, $\Mm$ is designed to guarantee that
\begin{align} \label{eq:sch2proofM}
\prod_{t=1}^T |(c_{t}-c^{\star}_{t})|^2 \ \dot\geq \ \theta^{2T}.
\end{align} This can be readily done for all dimensions $T$ by, for example, using the proper roots of unity as entries of a circulant $\Mm$ (cf.~\cite{BVR+:96}), which in turn allows for the above product - before normalization with $\theta$ - to take non-zero integer values.

In the post-whitened channel model at user $i = 1,2$, we have
\begin{align*} 
\bar{\bar{\yv}}^{(1)} &\defeq\diag( P^{-\delta^{(2)}_{1}/2}, \cdots, P^{-\delta^{(2)}_{T}/2} )\bar{\yv}^{(1)} \nonumber\\
  &=\diag( P^{-\delta^{(2)}_{1}/2}, \cdots, P^{-\delta^{(2)}_{T}/2} )\cv+  \bar{\zv}^{(1)}  \nonumber\\
	\bar{\bar{\yv}}^{(2)} &\defeq\diag( P^{-\delta^{(1)}_{1}/2}, \cdots, P^{-\delta^{(1)}_{T}/2} )\cv+  \bar{\zv}^{(2)}
\end{align*}
where, as we have stated, the noise $\bar{\zv}^{(i)}$ has finite power in the sense that
\beq \label{eq:noiseProb} Pr (||\bar{\zv}^{(i)}||^2 > P^{\epsilon})\rightarrow 0. \eeq

At the same time, after whitening at each user, the codeword distance for any two codewords $\cv,\cv^{\star}$, is lower bounded as
\begin{align}
 & ||\diag( P^{-\delta^{(i)}_{1}/2}, \cdots, P^{-\delta^{(i)}_{T}/2} ) (\cv - \cv^{\star}) ||^2  \nonumber\\
 & = \sum_{t=1}^{T}|P^{-\delta^{(i)}_{t}/2} (c_{t} - c^{\star}_{t}) |^2 \nonumber\\
 & \dotgeq \prod_{t=1}^{T}|P^{-\delta^{(i)}_{t}/2} (c_{t} - c^{\star}_{t}) |^{2/T}   \label{eq:sch2proofdis1}\\
& = P^{-\frac{1}{T}\sum_{t=1}^{T}\delta^{(i)}_{t}}  \prod_{t=1}^{T}|(c_{t} - c^{\star}_{t}) |^{2/T} \nonumber\\
 & \dotgeq P^{-\frac{1}{T}\sum_{t=1}^{T}\delta^{(i)}_{t}}  \theta^2  \label{eq:sch2proofdis2} \\
& \geq  P^{-\bar{\delta}^{\ast}} P^{\bar{\delta}^{\ast}+\epsilon} =  P^{\epsilon} \label{eq:sch2proofdis3}
\end{align}
for $i=1,2$, where \eqref{eq:sch2proofdis1} results from the arithmetic-mean geometric-mean inequality, \eqref{eq:sch2proofdis2} is due to~\eqref{eq:sch2proofM}, and where \eqref{eq:sch2proofdis3} uses the assumption that $\frac{1}{T}\sum_{t=1}^{T}\delta^{(i)}_{t} \leq \bar{\delta}^{\ast}$. Setting $\epsilon$ positive but vanishingly small, combined with \eqref{eq:noiseProb}, proves the result.
\end{proof}

At this point, we use the lattice code of the above lemma, to design the $T$-length vector $\cv$ transmitted during phase~$s$.  This encoding guarantees successful decoding of this vector, at both users, at a rate $R = r\log P -o(\log P)$, where $r=1-\bar{\alpha}^{(2)}$ for phase $S$, else $r= 1-\bar{\delta}$ ($\epsilon$ is set positive but vanishingly small, recall~\eqref{eq:sch1Combits}, \eqref{eq:sch1Combitss}). We note that for phase~$S$, user~$i=1,2$ can linearly transform their signal observations $\{ y_t^{(i)}\}_{t\in\Bc_S}$ (cf.~\eqref{eq:X1PhSy1},\eqref{eq:X1PhSy2}) to take the form in~\eqref{eq:latticech},\eqref{eq:latticech1}, while for phase~$s=1,2,\cdots,S-1$, user~$i=1,2$ can linearly transform their signal observations $\{ y^{(i)}_{t} - \bar{\check{\iota}}^{(i)}_{t} \}_{t\in\Bc_s}$ (after removing the interference $\bar{\check{\iota}}^{(i)}_{t}$, cf.~\eqref{eq:sch1ComIs},\eqref{eq:sch1y1},\eqref{eq:sch1y2}), again to take the form in~\eqref{eq:latticech},\eqref{eq:latticech1}.

Finally we note that the achievable rate is determined by the exponent average $\frac{1}{T}\sum_{t=1}^{T}\delta^{(i)}_{t}$ and not by the instantaneous exponents $\delta^{(i)}_{t}$.


\section{Appendix - Discussion on independence of estimation error and past estimates\label{sec:DiscussionAssumption}}
The assumption on independence of estimation error and past estimates, is consistent with a large family of channel models ranging from the fast fading channel (i.i.d in time), to the correlated channel as this was presented in \cite{YKGY:12d}\footnote{Note that our assumption is softer than the assumption in \cite{YKGY:12d} where $ \{ \{\hat{\bfh}_{t,t'},\hat{\bfg}_{t,t'}\}_{t'=1}^{t} , \bfh_{t}, \bfg_{t} \}_{t=1}^{t^{*}-1} \leftrightarrow \{\hat{\bfh}_{t^{*},t^{*}},\hat{\bfg}_{t^{*},t^{*}} \} \leftrightarrow \{\bfh_{t^{*}}, \bfg_{t^{*}}\} $ was assumed to be a Markov chain; an assumption which may not directly hold in block fading settings where for example, having $\bfh_{t^{*}-1} = \bfh_{t^{*}}$ (resp. $\bfg_{t^{*}-1} = \bfg_{t^{*}}$), breaks the chain $\{\bfh_{t^{*}-1}, \bfg_{t^{*}-1} \}\leftrightarrow \{\hat{\bfh}_{t^{*},t^{*}},\hat{\bfg}_{t^{*},t^{*}} \} \leftrightarrow \{\bfh_{t^{*}}, \bfg_{t^{*}}\}$ because, given $\bfh_{t^{*}-1} = \bfh_{t^{*}}$, the following conditional probability density functions hold $f_{ \hv_{t^*-1}, \hv_{t^*} | \hat{\hv}_{t^*}} = f_{ \hv_{t^*} | \hat{\hv}_{t^*}} \neq    f_{\hv_{t^*} | \hat{\hv}_{t^*}} f_{\hv_{t^*-1} | \hat{\hv}_{t^*}}$ as long as $\hat{\hv}_{t^*} \neq \hv_{t^*}$  (naturally  for two random variables $X_1$ and $X_2$, then $X_1|X_2$ cannot be independent of $X_1|X_2$ no matter what $X_1$ and $X_2$ are, unless $X_1 = X_2$).}, and even the quasi-static slow fading model where the CSIT estimates are successively refined over time. Successive CSIT refinement - as this is treated in \cite{Jindal:06m} - considers an incremental amount of quantization bits that progressively improve the CSIT estimates.
For example, focusing on the estimates of channel $\bfh_{1}$, the quality of this estimate would improve in time, with a successive refinement that would entail
\begin{align*}
\bfh_1 &= \hat{\bfh}_{1,1} +\tilde{\bfh}_{1,1}\\
&= \underbrace{\hat{\bfh}_{1,1} + \hat{\tilde{\bfh}}_{1,1,2}}_{\hat{\bfh}_{1,2} } + \tilde{\bfh}_{1,2} \\
&= \underbrace{\hat{\bfh}_{1,1} + \hat{\tilde{\bfh}}_{1,1,2} + \hat{\tilde{\bfh}}_{1,2,3} }_{\hat{\bfh}_{1,3} } + \tilde{\bfh}_{1,3} \\
& \quad \vdots
\end{align*}
where $$\tilde{\bfh}_{1,t'} \defeq \hat{\tilde{\bfh}}_{1,t',t''}+\tilde{\bfh}_{1,t''} $$ and where $\hat{\tilde{\bfh}}_{1,t',t''}$ denotes the estimate correction that happens between time $t'$ and $t''$.

Generalizing this to the estimate of any channel $\bfh_t$, and accepting that the estimate correction $ \hat{\tilde{\bfh}}_{t,t',t''}$ and estimate error $\tilde{\bfh}_{t,t''} $ are statistically independent, allows that the estimation error $\tilde{\bfh}_{t,t''} $ of $\bfh_{t}$ is independent of the previous and current estimates $\{\hat{\bfh}_{t,\tau} \}_{\tau\leq t''} $, which in turn allows for the aforementioned assumption to hold even for the block fading channel model.

As a side note, even though we consider the quantification of CSIT quality as in \eqref{eq:MMSEc}, we note that our results can be readily extended to the case where we estimate channel directions (phases), in which case we would simply consider $ \frac{1}{||\hv_t||}\hv_t=\hat{\hv}_t +\tilde{\hv}_t, \ \frac{1}{||\gv_t||}\gv_t=\hat{\gv}_t +\tilde{\gv}_t$ (cf.~\cite{XAJ:11b}).



\end{document}